# Adamantanes as white-light emitters: Controlling arrangement and functionality by external Coulomb forces

Jürgen Belz[1], Johannes Haust[1], Marius J. Müller[2], Kevin Eberheim[3], Sebastian Schwan[4], Saravanan Gowrisankar[5], Franziska Hüppe[1], Andreas Beyer[1], Peter R. Schreiner[5], Doreen Mollenhauer[4], Simone Sanna[3], Sangam Chatterjee[2], and Kerstin Volz[1*]

[1.] Department of Physics and Materials Science Center, Philipps-University Marburg, Hans-Meerwein Str. 6, 35032 Marburg, Germany.
[2.] Institute of Experimental Physics I and Center for Materials Research (ZfM), Justus-Liebig-University Giessen, Heinrich-Buff-Ring, 35392 Giessen, Germany.
[3] Institute of Theoretical Physics and Center for Materials Research (ZfM), Justus-Liebig-University Giessen, Heinrich-Buff-Ring, 35392 Giessen, Germany.
[4.] Institute of Physical Chemistry and Center for Materials Research (ZfM), Justus-Liebig-University Giessen, Heinrich-Buff-Ring, 35392 Giessen, Germany.
[5.] Institute of Organic Chemistry and Center for Materials Research (ZfM), Justus Liebig University, Heinrich-Buff-Ring 17, 35392 Giessen, Germany.

∗ Corresponding author: kerstin.volz@physik.uni-marburg.de

**Abstract**

Functionalized adamantane molecular cluster materials show highly transient nonlinear optical properties of currently unclear structural origin. Several interaction mechanisms in compounds comprising molecular clusters, their inter- and intramolecular interactions as well as the interplay of their electronic systems and vibrations of their backbone are viable concepts to explain these nonlinear optical properties. We show that transient Coulomb forces also have to be considered as they can lead to intramolecular structure transformations and intermolecular rearrangements in the crystal. Both strongly influence the nonlinear optical properties. Moreover, selective bromine functionalization can trigger a photochemical rearrangement of the molecules. The structure and chemical bonding within the compounds are investigated in dependence on the laser irradiation at different stages of their nonlinear emission by electron diffraction and electron energy loss spectroscopy. The transient structural and chemical states observed are benchmarked by similar observations during electron irradiation, which makes quantification of structural changes possible and allows the correlation with first principles calculations. The functionalization and its subsequent usage to exploit photochemical effects can either enhance two-photon absorption or facilitate white-light emission rather than second-harmonic generation.

Small changes in molecular structures such as exchanging a single atom in a bond with another can lead to drastic changes of their chemical and physical properties. A case in point is C–H-bond bromination that is a widely used tool to modify molecular properties of, e.g., pharmaceutically active compounds, but also in the area of agro- as well as materials chemistry.[1] Owing to the somewhat weaker C–Br bond (compared to C–F and C–Cl) and its photochemical lability, photochemistry often is exploited to achieve the desired structural changes that lead to changes in molecular and macroscopic properties.[1,2]

Here, we examine a model system for materials featuring highly nonlinear optical properties, which, however, can change depending on external stimuli. 1,3,5,7-Tetraphenyladamantane $AdPh_4$ exhibits white-light-generation (WLG) in the amorphous state and second harmonic generation (SHG), when crystalline.[3,4] The rigid correlation of optical studies, transmission electron microscopy (TEM), and results from first principles calculations reveals that Coulomb effects lead to structural modifications in the material, which can be exploited to manipulate the functional material in white-light emitting devices. In this work we utilize a partially brominated set of phenyladamantanes of the general formula $AdBr_xPh_{4-x}$ ($x \in [0, \ldots, 4]$) to exploit the light-stimulated transformation of the crystalline to the amorphous state in a changed intermolecular arrangement upon altered intramolecular structure. We thereby exploit the photochemical lability of the C–Br bonds that leads to a strong transient nature of its molecules' intramolecular structure upon energy deposition. We tailor this stabilizing effect to derive guiding principles for future WLG materials, which is only possible by the established correlation between molecule's intramolecular structure and its impact on intermolecular arrangement resulting in changed structure and bonding with its nonlinear optical properties. Hereto, we correlate time-dependent TEM diffraction as well as optical spectroscopy with *ab initio* first principles calculations on functionalized adamantanes.

**Results and Discussion**

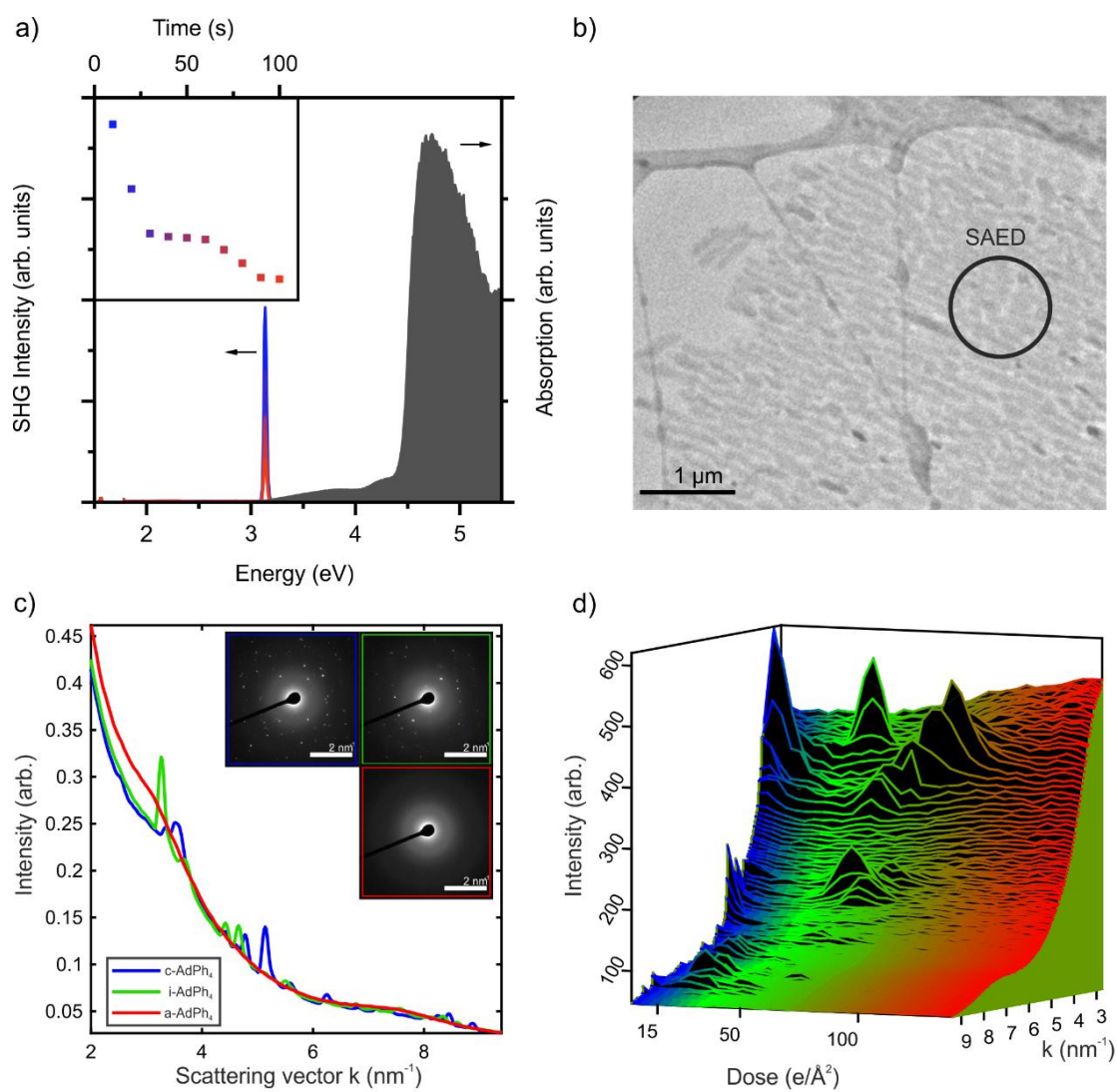

*Figure 1: (a) Linear absorption (dark grey) and nonlinear optical response for pristine $c\text{-AdPh}_4$. The SHG intensity decreases with time (blue to red curves). The temporal evolution of the emission intensity is illustrated as inset. (b) TEM image of $c\text{-AdPh}_4$ supported by a lacey carbon film and embedded in epoxy resin (no contrast). The location as well as the diameter of a selected area electron diffraction (SAED) aperture is indicated exemplarily. (c) The diffraction patterns of the exposure series show the pristine (blue), intermediate (green) and amorphous state (red), achieved by electron irradiation. The radial scattering distribution is shown with the respective colors below the diffraction patterns. Crystalline diffraction is changing and finally vanishing. In addition, an increase of the continuous diffuse scattering background can be observed. (d) Full exposure series as a waterfall diagram. With increasing dose, the initially crystalline diffraction pattern changes into an intermediate state and finally fades into a non-diffracting diffuse scattering from which the intermolecular atomic arrangement will be determined.*

One technologically highly relevant example for molecular materials showing transient properties is the nonlinear response of 1,3,5-7-tetraphenyladamantane ($AdPh_4$). The amorphous form of this compound shows WLG for ultrafast near-infrared excitation and second harmonic generation if crystalline. [3] However, the efficiency of the second harmonic generation decreases for intense laser irradiation under high-vacuum environments, see Figure 1a. The photon energies of the exiting laser are chosen at 1.51 eV. This is below the onset of two-photon absorption, *i.e.*, less than half the optical gap at 4.47 eV derived by Tauc's method, [5,6] also see supporting information. This hints at a change of the structure of the compound itself.

In the following, we investigate the morphological and chemical changes in detail starting with the case of $AdPh_4$. Figure 1b shows a transmission electron microscope (TEM) overview image of c-$AdPh_4$ crystallites that are supported by lacey carbon film and surrounded by epoxy (no contrast). Transient changes in the crystal structure can be observed. Amorphization of the initially single-crystalline regions can be concluded from Figure 1c showing the change of the diffraction patterns (DP) during electron irradiation under dose-controlled conditions. The pristine crystalline DP (Figure 1c, blue) changes considerably after an irradiation of about 16 $e/Å^2$ (Figure 1c, green), followed by a total absence of crystalline diffraction at about 100 $e/Å^2$ (Figure 1c, red). In this final amorphous state, the background intensity changes and forms diffuse rings where diffraction occurred before. This suggests disordering is taking place. The existence of the rings also indicates the existence of non-random structures [7] as it will be elaborated in more detail later in this manuscript and in the supporting information. The dose series of these changes is depicted as the radial sum in a waterfall plot (Figure 1d). Due to the very fast initial spot-intensity decay (Figure 1d, blue shaded area), we use additional ultra-low dose (ULD) measurements with exposures of 0.33 $e/Å^2$ per DP to investigate the initial change more quantitatively. This ULD measurement reveals that the crystalline DP shows an additional, but comparatively minute change within the first frames already at very low total doses (< 1 $e/Å^2$). This initial effect might be associated with charging and subsequent rotations and translations of either the crystallites or/and molecular building blocks in the crystalline framework. The rotation of the phenyl-groups as supported by density function theory (DFT) calculations, presented later in the text (Figure 5), would be in accordance with this observation, since the rotation barrier for single molecules is sufficiently low.

In order to investigate the structural and chemical changes in detail, we employ TEM electron energy loss spectroscopy (EELS). In particular we investigate potential changes in the molecule´s intramolecular structure, depending on the treatment (pristine vs. electron-beam-amorphized vs. laser-amorphized). Thereto, we use the energy loss at the carbon K-shell that shows a prominent ionization edge. The energy loss near edge structure (ELNES) of such a so-called core-loss EELS contains information about the bonding environment of carbon atoms which results in distinct features that are relatable to bond types and the chemical environments.[8] One of the most prominent features is the 1s-π* C=C peak at about 285 eV energy loss, where *s*-orbital electrons of *sp²*-hybridized carbon atoms can transition into the antibonding π orbital (π*). The transition into the antibonding σ orbital (σ*) 1$s$-σ* C-C is typically located at around 293 eV, but can have different values depending on the bonding type and atom.[9]

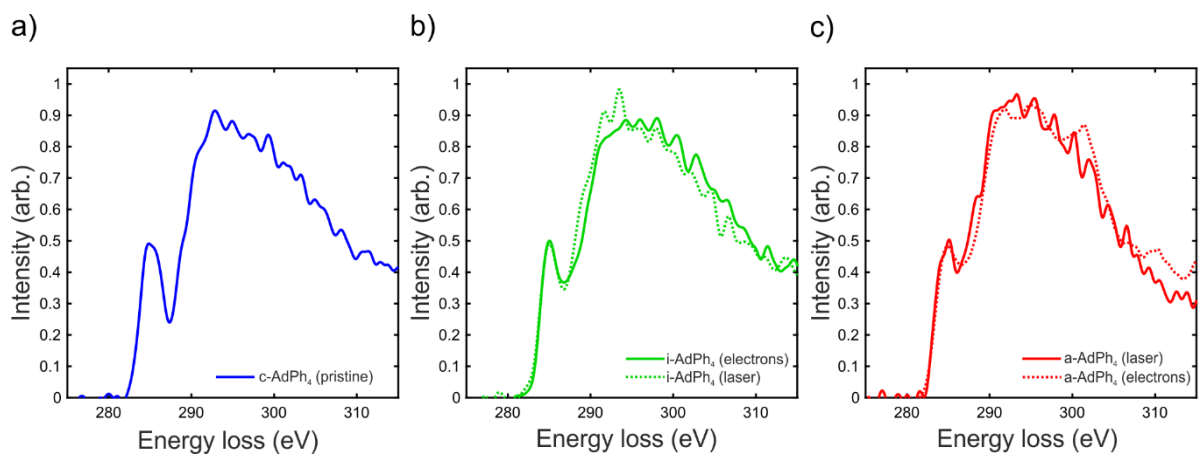

*Figure 2: Carbon K-shell ionization edges of different $AdPh_4$ states show a varying amount of oriented sp²-bonds indicated by the peak located at 285 eV. (a) The pristine crystalline specimen shows the most prominent peak associated with the excitation from 1s-π* C=C carbon bonds. (b) For the laser- and electron-irradiated specimen with medium exposures we find a decrease of this signal that is very similar for both types of irradiation. (c) In the final amorphous state, the peak is even less pronounced, and both irradiation types lead to similar final states.*

Figure 2a shows the resulting carbon K-shell ionization spectrum from pristine $c\text{-}AdPh_4$. It can be seen that in contrast to the irradiated specimen shown in Figure 2b and Figure 2c the *sp²*-peak is very pronounced. The EELS from intermediate dose exposures shown in Figure 2b demonstrate a very similar impact of electron- and laser-irradiation. A further loss of *sp²*-signal can be seen for $a\text{-}AdPh_4$. This trend can be explained by two mechanisms: the more obvious mechanism is the loss of *sp²*-hybrization which implies the destruction of phenyl-groups. The other explanation is connected to the alignment of the π-electron system with respect to the impinging beam. In a crystalline specimen a distinct alignment exists, whereas amorphous materials show random alignment. This effect is known to complicate the bond type determination.[8] Nearly free rotation of the phenyl-groups, as also corroborated later by DFT-calculations (Figure 5), strongly supports the latter as dominant mechanism for our observations.

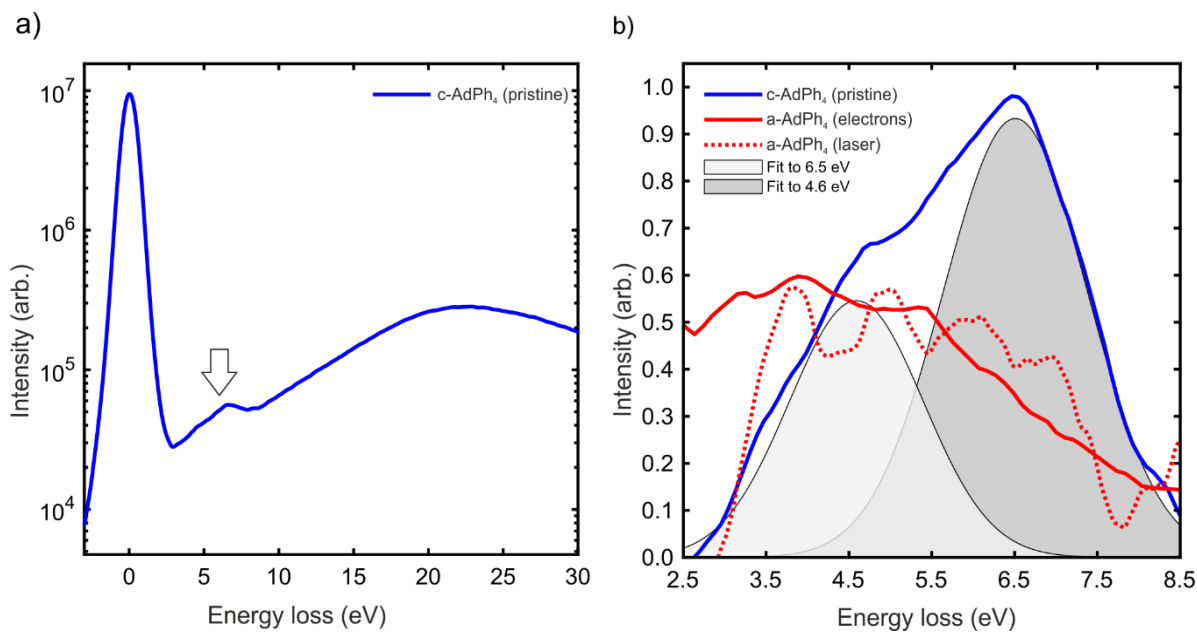

*Figure 3: (a) Low-loss EELS showing a prominent feature associated to the sp²-hybridization at around 6.5 eV (arrow) for $c\text{-}AdPh_4$. (b). The laser- and electron-irradiated specimen show a very different low-loss spectrum (red), which cannot be decomposed fully into peaks located at 4.6 eV and 6.5 eV. Those peaks correlate with the adamantane absorption (4.6 eV) and the single-electron transitions associated to the π electron system (6.5 eV). The change of peak ratios suggests a rotation or loss of phenyl-groups.*

These findings are confirmed by the investigation of the low-loss EELS (Figure 3). In these spectra, a prominent feature can be seen at around 6.5 eV only for the crystalline state (overview shown for the pristine compound in Figure 3a). This peak is known to be associated with a single-electron transition of π-electrons.[10] In both amorphized specimen, this regime is broadened and the prominent single-electron transition is reduced, suggesting the aforementioned rotation of the phenyl groups upon irradiation. The reduction of the characteristic peak at 6.5 eV is illustrated in Figure 3b. The low-loss feature can be deconvolved into two distinct signals located around 4.6 eV and 6.5 eV, respectively, for the case of $c\text{-}AdPh_4$ only. The ratio of these spectral features is very different for the irradiated specimen, and a multi-Gaussian model requires at least three peaks to fit the data accurately. Nonetheless, we can connect the peak centered at 4.6 eV to the optical absorption (Figure 1a and Figure 6) of the phenylated adamantanes under investigation. A detailed study of the transitions, which contribute to the low-loss region, will be subject to future investigations.

It is an important finding that upon amorphization of $AdPh_4$ the phenyl-groups rotate and that the entire molecule stays intact, yet rearranges. This is confirmed by evaluating the *differential electron scattering intensity* (*dI*) of pristine as well as electron- and laser-amorphized $AdPh_4$ (Figure 4), the raw diffraction patterns of these samples are shown in the supporting information (*cf.* Figure S3). The scattering shows characteristic deviations from the average atomic form factor of the constituents, indicating deviations from random atom positions, and hence distinct correlations of atom sites.[11] These features underline the existence of molecules in random orientations. We carry out molecular dynamics (MD) simulations by using the OPLS-AA force field of a large number of molecules (1728) in a super cell to model an amorphous arrangement (*cf.* supporting information) followed by electron scattering simulations using the in-house developed STEMsalabim.[12] Both measurements of laser- as well as electron-amorphized a-$AdPh_4$ shown in Figure 4a yield the same qualitative behavior and all oscillations are observed at nearly identical positions as found in the simulation. Figure 4b shows the pair distribution function (PDF) derived from the scattering data. It can be seen that all materials follow the general structure of $AdPh_4$. Especially the nearest-neighbor (NN) and second nearest-neighbor (2NN) peaks are clearly located at the anticipated locations in good agreement with the scattering simulations. These data clearly underline that the intramolecular structure of the molecule is conserved for electron- as well as laser-amorphized samples. Shoulders at the positions of the 1NN and 2NN associated with the phenyl-related C-C distance also suggest that the entire molecule is still intact and that the phenyl-groups just rotate (the rotational barrier of the phenyl-groups for a single molecule have been calculated to be 4 kJ mol$^{-1}$ (41 meV) at the B3LYP-D3(BJ)/cc-pVDZ level of theory.

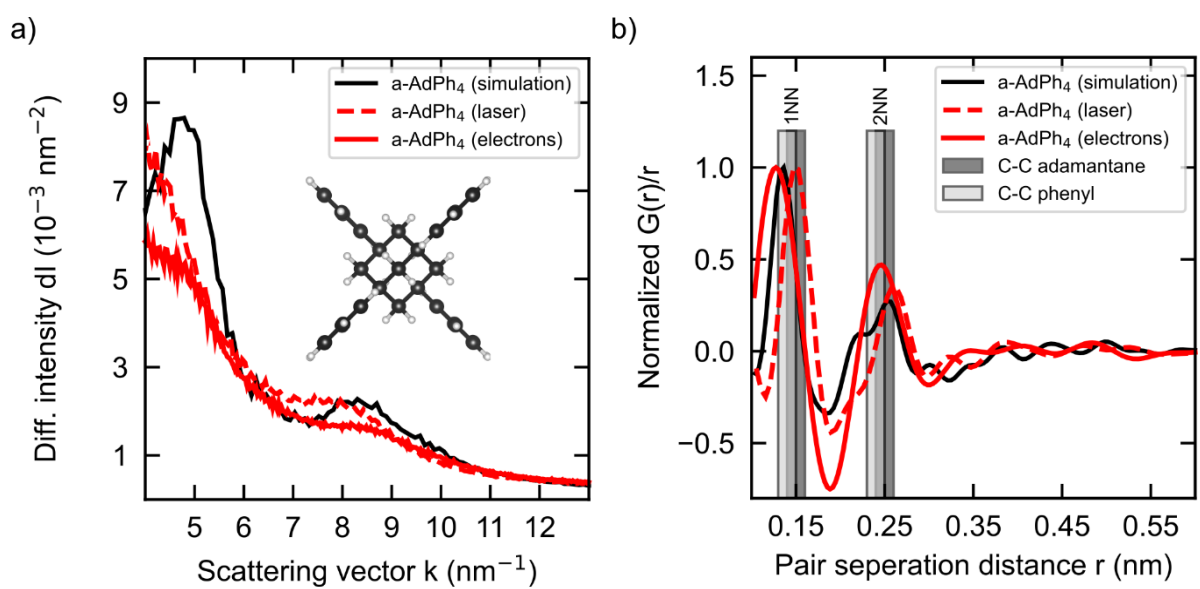

*Figure 4: (a) Radial scattering distribution comparing the different amorphous states of $AdPh_4$ (laser-amorphous, electron-amorphous) with an electron scattering simulation assuming a random arrangement of $AdPh_4$ molecules obtained as result of MD simulations. (b) The PDFs are derived from the scattering data and show the oscialltions at the characteristic nearest neighbor distances of $AdPh_4$.*

In the following, we focus on the brominated derivatives $AdBr_xPh_{4-x}$ ($x \in [0,...,4]$). With this family of molecules, we aim to enhance the structural reorganization of the compound by systematically substituting the phenyl-groups by bromine. The high photochemical reactivity of bromine[2] makes this partial halogenation a promising candidate to facilitate the structural transformation from a polycrystalline into an amorphous state using electromagnetic forces. In order to investigate this, the impact of electron irradiation on the molecular structure of $AdBr_xPh_{4-x}$ single clusters is modeled by means of atomistic *ab-initio* calculations. During irradiation, different processes may occur, leading to both electron accumulation as well as to electron ejection. Thus, both electron excess and electron deficiency are considered in our models. As the exact amount of electronic charge transferred to the single molecules is difficult to quantify, we consider the effect of different integer numbers of excess electrons on single $AdBr_xPh_{4-x}$ molecules. While this approach is justified by the fact that in insulating molecular compounds single electrons are not delocalized over many molecular clusters, it is intended to model the molecular response to excess charges only qualitatively.

Our calculations reveal that electron depletion has only a minor effect on the whole $AdBr_xPh_{4-x}$ series and will not be discussed in detail. On the contrary, excess electrons lead to drastic structural modifications, which are schematically represented in Figure 5.

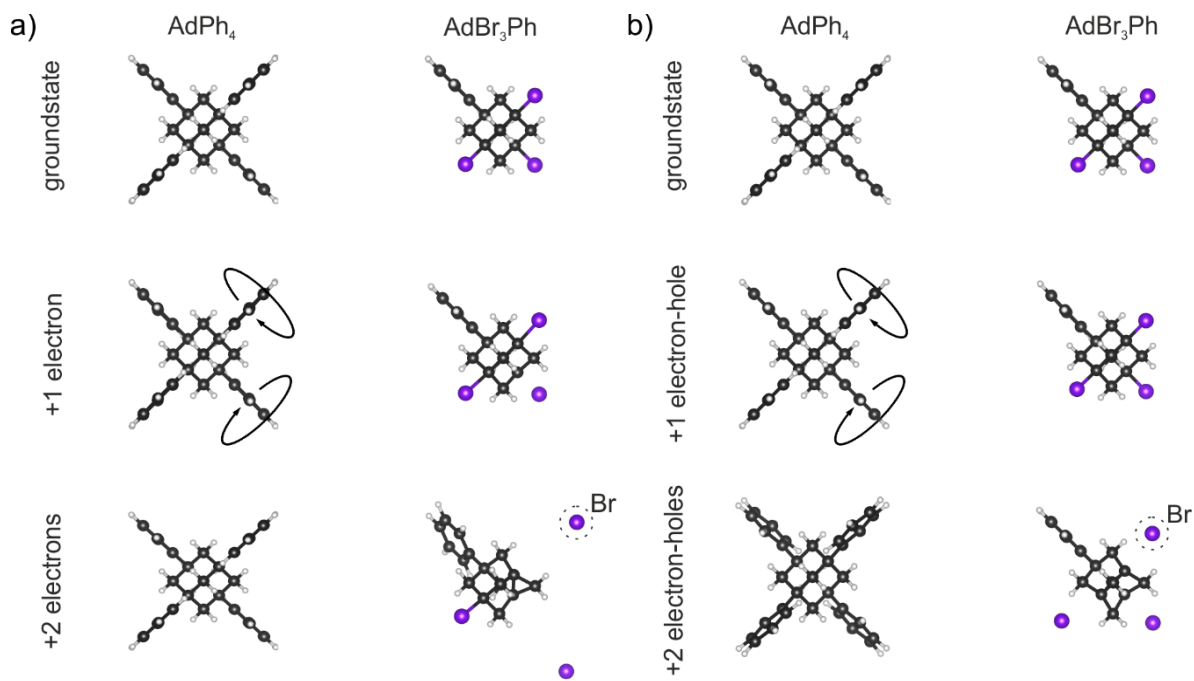

*Figure 5: Structural models of the $AdBr_xPh_{4-x}$ series in the neutral charge state (first row) as well as upon electron doping (a) and after formation of one or more electron-hole pairs (b). For clarity, the structures are shown for $AdPh_4$ and $AdBr_3Ph$, structures for all clusters can be found in the supporting information. The cluster structure is calculated within DFT-PBE and a plane-wave basis as specified in Methods.*

$AdPh_4$ clusters (Figure 5a, first panel) are robust and do not undergo severe structural modifications upon doping with up to four electrons per clusters. With an increasing number of electrons, the phenyl-rings are rotated with respect to their original orientation, a finding that can also explain the changes in the DPs already observed for minuscule electron doses (*cf.* Figure 1) as well as the core-loss and low-loss EELS observations (*cf.* Figure 2 & 3).

Br-substituted $AdBr_xPh_{4-x}$ clusters behave very differently instead. As shown exemplarily for $AdBr_3Ph$ in Figure 5a, drastic structural modifications are expected upon doping (structural details and the full $AdBr_xPh_{4-x}$ series are given in the supporting information). Two different regimes can be distinguished. For a low fraction of excess electrons, the structural modifications are limited to a torsion of the phenyl-rings. In this regime, electron doping is expected to induce disorder in the compound by destroying the long-range order of the molecular compounds, without significantly impacting the geometry of the single cluster. In particular, the cluster core remains intact. For higher numbers of excess electrons, the clusters undergo large structural deformations, including detachment of the Br-substituent and partially involving the cluster core, which becomes reminiscent of a 1,3-dehydroadamantane core structure that is accessible, *e.g.*, via Wurtz coupling.[13] Thus, true chemical reactions occur upon high electron doping.

This behavior can be understood from the electronic structures of the investigated clusters. Atomic Br (electronic configuration [Ar]$3d^{10}4s^24p^5$) is highly electro-negative (2.96 Pauling units). Correspondingly, the (antibonding) LUMO orbital of $Br_xPh_{4-x}Ad$ molecules is a C–Br hybrid, strongly localized at the Br atom(s), as shown in the supporting information (Figure S6). When electron doping exceeds a certain value, dictated by the degree of hybridization, the Br-substituent may reach a very stable closed shell configuration, and detach as an anion from the cluster core. Subsequently, the cluster geometry rearranges, partially involving the cluster core as shown in the last row of Figure 5a (and in the supporting information, Figure S7). Summarizing, electron irradiation heavily impacts the crystal order, either by destroying the long-range order of the molecular crystals (low dose electron radiation) or by chemical modifications of the clusters (higher dose electron irradiation).

As the occupation of Br-related unoccupied states leads to the major modifications of the molecular geometry discussed above, it is reasonable to expect a similar effect upon laser irradiation. The latter is indeed expected to lead to the occupation of Br related LUMO states upon creation of electron-hole pairs. To address this question, one or more e-h pairs have been created and frozen in the isolated clusters of the $AdBr_xPh_{4-x}$ family. The corresponding structural modifications as calculated from DFT-PBE in a plane-wave basis are shown exemplarily for $AdPh_4$ and $AdBr_3Ph$ in Figure 5b (full series in the supporting information, Figure S8). Laser irradiation has an impact on the molecular cluster which is rather similar to that of electron irradiation, suggesting that indeed the occupation of the Br-related, unoccupied states is responsible for the structural modifications. A single e-h pair leads again to the rotation of the phenyl rings. A higher number of Coulomb pairs leads to the detachment of the Br-radicals and more profound deformations. This ultimately leads also to a 1,3-didehydro-diphenyladamantane core moiety in the case of *AdPh$_2$Br$_2$* (*vide infra*).

Concluding, atomistic calculations suggest that sufficient exposure to both, free electrons or laser excitation should lead to amorphization of crystal from the $AdBr_xPh_{4-x}$ family-of-compounds through various processes. These range from changing the relative orientation of the clusters for low irradiation conditions to actual change of their form or even composition for stronger irradiation conditions.

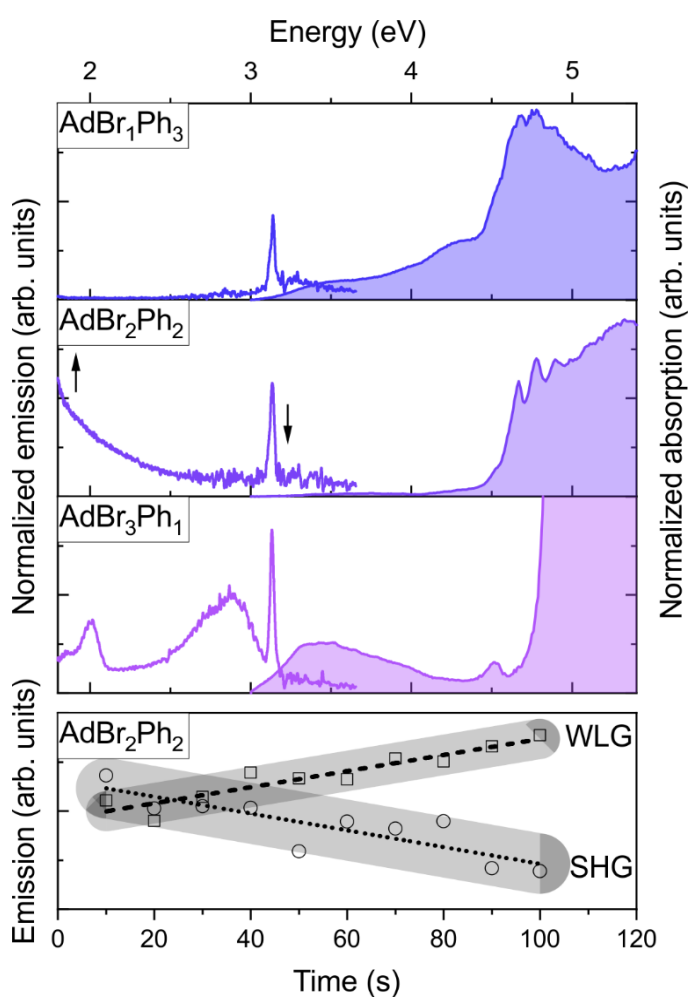

*Figure 6: Linear absorption data of the $AdBr_xPh_{4-x}$ family-of-compounds. All show a pronounced absorption edge around 4.5 eV. AdBrPh$_3$ and AdBr$_3$Ph show additional, pronounced features towards visible photon energies (shaded areas). The nonlinear optcal response of the compounds is also shown as solid lines. The bottom panel quantifies the changes of the second-harmonic (decrease, circles) and continuum emission (increase, squares) as function of the exposure time for $AdBr_2Ph_2$.*

In order to investigate the effect of a different number of C–Br-bonds on the stability and the optical properties as well as their transient behavior, we synthesized the full family of $AdBr_xPh_{4-x}$ and investigate their optical response systematically. Figure 6 summarizes the optical response of the compounds under investigation. The linear absorption inferred from reflection data under ambient conditions are plotted as shaded areas. All compounds show a steep increase in absorption around 4.5 eV. These are associated with dipole-allowed direct absorption edges according to Tauc's method[5,6] (*cf.* Figure S1). These transitions are also found in the low-loss EELS data of the respective compounds, irrespective of irradiation (*cf.* Figures 3b and 8b).

All pristine compounds show SHG as seen from the solid lines in Figure 6. This optical nonlinearity is characteristic for crystalline samples with lack of inversion symmetry. Here, one has to consider that polycrystallinity or the existence of crystalline inclusions in an amorphous matrix can break the inherent symmetry of bulk single crystals.[14] In addition, the change between different dielectric environments at surfaces or hetero-interfaces is inherently associated with a second-order nonlinearity that can cause second-harmonic generation. Next, we consider the dependence of the observed nonlinear optical effect for $AdBrPh_3$. Here, we observe both, SHG as well as two emission bands centered around 2.85 and 2.0 eV, respectively. The latter are attributed to spontaneous emission following two-photon absorption into the low-energy absorption features. The two-photon absorption is related to the imaginary part of the second-order nonlinear susceptibility and typically outweighs SHG, which is related to the real part of the second-order nonlinear susceptibility. Consequently, SHG in this compound is generally weak. The spontaneous emission feature decreases with irradiation exposure time, similar to SHG. To understand this, we have to consider that the C–Br-bond is expected to dissociate under photoexcitation as predicted by theory. Consequently, the molecules are easily photochemically transformed, hence change their intramolecular structure, and should in turn alter their intermolecular arrangement to render an – most likely – amorphous material with residual *sp$^2$*-hybridized substituents.[15] For $AdBrPh_3$, any two-photon excitations apparently decay nonradiatively, as virtually no signatures of spontaneous emission are observed. The even weaker SHG also decreases with increasing irradiation time, suggesting a similar photochemical mechanism to $AdBr_3Ph$.

Next, we turn to $AdBr_2Ph_2$ that demonstrates a quantitatively very different behavior. The compound shows SHG for pristine material upon weak irradiation. For more intense irradiation, the conversion efficiency into SHG decreases with irradiation time and the compound starts to show a spectrally broad emission at lower energies. Its efficiency increases as the SHG emission decreases (Figure 6, bottom panel). This infers that $AdBr_2Ph_2$ photochemically transforms: its crystalline structure appears to vanish, reducing SHG conversion. Decreasing the incident photon flux after photochemical transformation retains the nonlinear response with decreasing conversion efficiency. Alternatively, the broadband emission could originate from thermally stimulating the material. However, observing significant WLG with significant spectral components in the visible requires temperatures well above the thermal dissociation threshold of typically about 250 K even when assuming perfect emissivity. Fitting the calibrated spectra with black-body radiation and assuming perfect emissivity yields an initial temperature of about 2600 K which gradually decreases to about 1600 K (*cf.* Figure S2).

All these observations agree with the following scenario: The unsubstituted $AdPh_4$ stays intact as molecule, however, the *$sp^2$*-hybridized phenyl-ligands rotate already at low irradiation doses, resulting in a structural transformation as described above. Despite the amorphousness of the material, WLG is not observed as the phenyl-groups continuously rotate. The substitution of a single phenyl-ring with bromine does not affect the photochemistry significant enough to drastically change this. Replacing two phenyl-rings with bromine appears to optimize the amount of residual *$sp^2$*-hybridized phenyl-groups retained to allow for continuum emission. Molecular compounds featuring larger two-photon absorptions (at around 3 eV for a laser exciting at 1.51 eV photon energy) could easily thermally decompose.

In analogy to the investigations of the purely organic *AdPh₄*, we also investigated bromophenyladamantane derivatives by electron microscopy to quantify the structural modifications expected from photochemistry after both, laser as well as electron irradiation. We show $AdBr_3Ph$ as it is the most photochemically sensitive compound in this series. $AdBr_3Ph$ can be made amorphous already for ULD (< 1 $e/Å^2$) measurements, hence we call the pristine state $a\text{-}AdBr_3Ph$ (pristine). Figure 7a shows the *differential scattering intensity* (*dI*) of the laser exposed and electron irradiated states.

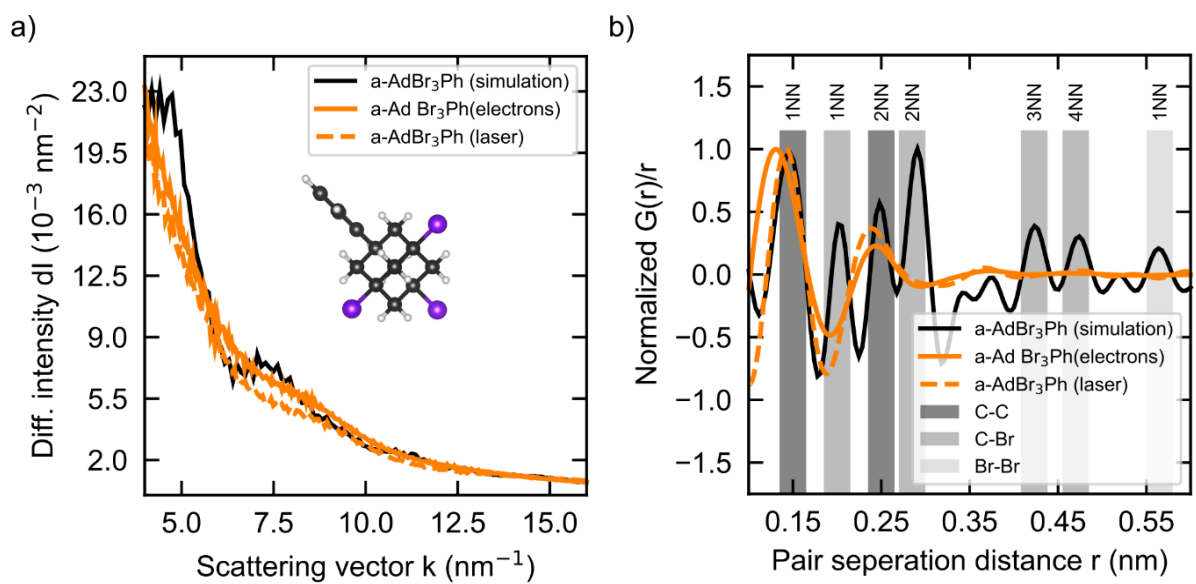

*Figure 7: (a) Scattering distribution of different $AdBr_3Ph$ states. The structural peaks are found on every curve with different intensity. (b) The PDFs derived from these scattering curves show the characteristic oscillations that can be attributed to the adamantane. Both irradiated curves show the same features indicating the same structure. The simulation (black) assuming intact, though randomly oriented $AdBr_3Ph$ discussed in the text.*

Both curves show characteristic deviations from the average atomic form factor of their constituents, indicating correlations of atom sites. Super cells derived from MD simulations of a large number of molecules in an amorphous arrangement serve as input for the scattering simulations. The derived PDF shows distinct differences between the experimental data and the corresponding simulation of intact molecules (Figure 7b). While the PDFs of both irradiated compounds are extremely similar, (dashed and solid orange curves), neither coincides with the curve derived from theory for the brominated adamantane (solid black line). The only common peaks are associated to the nearest (NN) and second nearest (2NN) neighbor distances of carbon atoms (C-C) that predominantly originate from the adamantane core and the remaining phenyl ring. This already indicates that irradiated $AdBr_3Ph$ still retains the adamantane core, similar to the $AdPh_4$ (Figure 4). However, any correlations related to bromine atoms are missing. Consequently, these specimens do either no longer contain brominated adamantanes or in very limited amounts. As suggested by DFT (Figure 5a), we can anticipate that Br is still bound to the adamantane in the pristine state. Notably, the experimental scattering curves of the laser irradiated compound can only be reproduced without bromine atoms within the compound while the data for electron irradiated compounds require the inclusion of Br to achieve a similar goodness-of-fit. These findings are consistent with electron energy loss spectroscopy (EELS) of the bromine M-shell ionization edge, as well as energy dispersive X-ray spectroscopy (EDS) (*cf.* Figure S5).

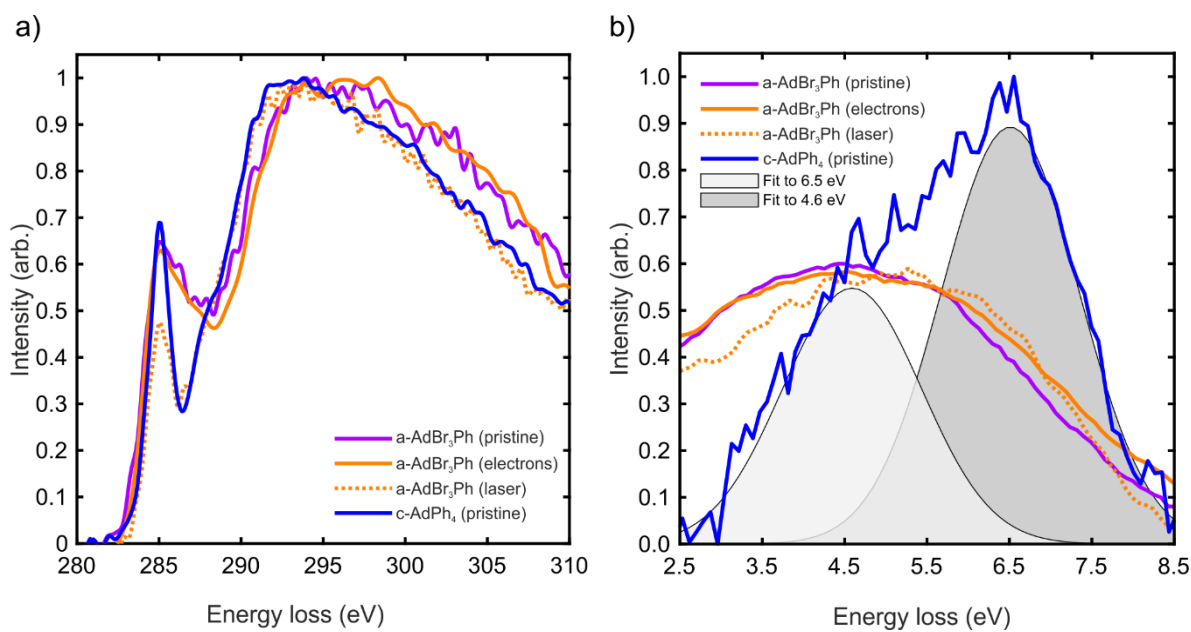

*Figure 8: (a) The EELS at the carbon K-shell ionization edge shows two distinct shapes for different states of $AdBr_3Ph$. The pristine and electron-irradiated curves show the same appearance and are blue-shifted by about 2 eV with regard to the $AdPh_4$ molecule. The laser-irradiated $AdBr_3Ph$ follows the curve of $AdPh_4$ almost perfectly with a reduced $sp^2$-signal at 285 eV. (b) The low loss region shows a distinct difference between aforementioned groups. This region is for all brominated materials qualitatively similar and shows a reduced signal for the 6.5 eV signature compared to the crystalline $AdPh_4$ (cf. Figure 3b). The laser-irradiated specimen shows a small tendency towards the peak ratio observed for $c\text{-}AdPh_4$.*

With this, we can conclude that both, $a\text{-}AdBr_3Ph\,(laser)$ and $a\text{-}AdBr_3Ph\,(electrons)$ have a common intramolecular structure, *i.e.*, the adamantane core without any or only minor deviations from its PDF fingerprint. The Br, which is still incorporated in the electron-irradiated specimen, however, mostly in a non-bonded form, is absent in the laser-irradiated specimen, as for these de-bromination took place prior to TEM sample preparation and hence Br-diffusion or formation of volatile Br-compounds results in desorption of the bromine from the material.

The de-bromination of $AdBr_3Ph$ and the retaining of the phenyl-group at the adamantane core upon laser irradiation is further corroborated by EELS. Figure 8 shows the EELS data of the pristine as well as laser- and electron-amorphized $AdBr_3Ph$ and $c\text{-}AdPh_4\,(pristine)$ for reference. The pristine $a\text{-}AdBr_3Ph$ and the $a\text{-}AdBr_3Ph\,(electrons)$ show an almost identical carbon K-shell ionization edge (Figure 8a), which significantly deviates from the data obtained for $a\text{-}AdBr_3Ph\,(laser)$. The latter almost exactly coincides with the shape of $c\text{-}AdPh_4$. The shift of the main K-shell ionization edge towards higher energies for the electron-irradiated material, which still contains bromine, is about 2 eV [1], which is consistent with DFT calculations of 1$s$-σ* C-C bond energies in Br-containing $AdBr_3Ph$. This underlines that Br is still present in the electron-irradiated samples and can influence the carbon bond energies. We suspect the majority of bromine to be unbound to the adamantanes as suggested by the PDF (Figure 7b).

Compared to the $c$-$AdPh_4$, the laser-irradiated $AdBr_3Ph$ shows a significantly reduced intensity for the peak at 285 eV, which is associated to the *sp²*-hybridized carbon atoms, in accordance with the reduced number of phenyl-groups, which, however, are still intact. The signal for the electron-irradiated $AdBr_3Ph$ in this energetic region is broadened towards higher energy losses, which could be connected to the electronic interference of intercalated bromine in the vicinity of the phenyl-groups.

The low loss data (Figure 8b) of the various $a$-$AdBr_3Ph$ compounds further clarify the structure. All $a$-$AdBr_3Ph$ exhibit quite similar signals, which are distinctly different from the $c$-$AdPh_4$ shown for reference. The single-electron transition (π→π*) of *sp²*-hybridized carbon at around 6.5 eV [10] is reduced for $a$-$AdBr_3Ph$ compared to $AdPh_4$, corroborating the reduced Ph/Ad ratio in these compounds. Multi-Gaussian decomposition reveals that the ratio of the peaks at around 4.6 and 6.5 eV stays virtually constant for all $a$-$AdBr_3Ph$ irrespective of irradiation treatment, again underlining that the Ph-adamantane bonding in all $a$-$AdBr_3Ph$ compounds is intact.

---

[1] *The electronic states related to the C-Ph bond in $AdPh_4$ clusters are characterized by an energy of -269.43 eV. While the energy related to the C-Ph bond in $AdBr_xPh_{4-x}$ clusters is very similar, the electronic states related to the C-Br bond are characterized by a lower energy. In the case of $AdBr_3Ph_1$, e.g., it amounts to -271.26 eV.*

**Summary**

We quantify the effect of transient Coulomb forces by either electron- or laser-irradiation on the structure and nonlinear optical properties of the organic tetra-phenyl-adamantane ($AdPh_4$) as well as the brominated derivates $AdBr_xPh_{4-x}$ to explain structural transformations and resulting changes in the optical response. For $AdPh_4$, we observe the same behavior irrespective of the type of irradiation: its initial crystallinity is destroyed at very small doses by rotation of the phenyl-groups. This causes rotation/movement of the adamantane cores, resulting in an amorphous intermolecular arrangement, however, preserving the intramolecular structure, *i.e.*, the phenyl-rings as well as the adamantane nature of the core. For the brominated derivates, the Coulomb forces result in an electronic excitation and, hence, in a changed intramolecular structure by de-bromination. Similarly, the molecules lose their crystalline intermolecular order for electron as well as laser irradiation and undergo transition from polycrystalline order to a disordered, *i.e.*, amorphous state. All these observations are in line with DFT computations and result in highly transiently changing nonlinear optical properties: depending on the number of C–Br-bonds, a transition from SHG to a broadband continuum emission is observed. These findings are indeed highly technologically relevant, as the optical properties can be deliberately modified, even at different length scales to generate structured white-light emitters.

**Methods**

**Organic Synthesis**

All chemicals and solvents were purchased at the highest commercial grade from Acros Organics (Adamantane, $CBr_4$, $Bu_4NBr$), Sigma-Aldrich ($AlCl_3$, $AlBr_3$, Benzene, MeOH), Alfa Aesar (Fluorobenzene, *tert*-BuBr) and used as received. Prior to use, the glassware was oven-dried at 80 °C overnight, allowed to cool to room temperature, and purged with argon. All yields refer to yields of isolated pure products. All products were identified using nuclear magnetic resonance (NMR) analysis. $^1$H NMR and $^{13}$C NMR spectra were recorded in $CDCl_3$ on Bruker spectrometers at 400 ($^1$H NMR) and 100 MHz ($^{13}$C NMR). All shifts are reported in parts per million (ppm) relative to residual $CHCl_3$ peak or trimethylsilane (TMS) as internal standard. All coupling constants (J) are reported in Hertz (Hz). The detailed experimental procedures and analytical data are summarized in the supporting Information.

**Optical characterization**

The linear optical responses of the compounds are derived from reflectance spectroscopy data using a deuterium light source in combination with UV compatible optics in a home-built microabsorption/confocal reflection setup featuring *f/1* optics. The instrument response covers 5.5 eV to 1.25 eV and yields sub-µm spatial resolution. A deep-UV optimized plane Al mirror is used as reference for the absorption data. The samples were measured under ambient condition.

The nonlinear optical response is measured in a tailored micro-photoluminescence setup. The samples are mounted in vacuum on the heat exchanger of a helium-flow cryostat mounted on a *xy*-nanopositioning stage with 20 nm resolution in closed-loop mode (Smaract GmbH). The compounds are irradiated using 100-fs near infrared pulses centered at 800 nm generated by a 78 MHz repetition rate Ti:sapphire laser oscillator using a 60 x, 0.65 NA infinity-corrected, cover-glass-corrected microscope objective. The nonlinear response is recorded in backscattering geometry and focused onto the entrance slit of using a 28 cm Czerny-Turner imaging spectrograph (Andor Kymera 328i) using a superachromatic tube lens (B-Halle Nachfolgegesellschaft). The spectrally dispersed light is recorded using a thermoelectrically cooled UV optimized back-illuminated deep depletion CCD camera (Andor 420BUDD) The driving laser is suppressed using short-pass filters (Thorlabs FESH700).

The spectral response is calibrated using a traceable tungsten-halogen source. The calibration spectra providing the instrument response are given in the supporting information.

**Transmission electron microscopy and related specimen preparation**

Electron transparent specimens are prepared by room temperature wet-cutting ultramicrotomy (UMT) using a Leica ULTRACUT UCT. Hereto, the material is embedded in a suitable araldite epoxy and cured at slightly elevated temperatures of about 80°C. The slice thickness is set to be 50 nm. The UMT cuts are lifted from the water and transferred to lacey carbon TEM support mesh grids. Before investigation, the specimens are dried and degassed in vacuum.

Conventional broad beam illumination transmission electron microscopy (TEM) as well as scanning TEM (STEM) are used to quantify the structure of the compounds. We carry out diffraction experiments using a JEOL JEM-3010 TEM with parallel illumination operating at 300 kV. For the diffraction experiments we use ultra-low doses (ULD) down to about 0.2 $e/Å^2$ per exposure with a careful minimum dose specimen approach. The microscope is equipped with a sensitive and fast TVIPS X416F-ES 4k camera. Measurements are carried out at room temperature. In order to reduce the effect of contamination, we use an adjacent region of no interest to polymerize adsorbed hydrocarbons by electron flooding with the maximum available dose rate for about 30 minutes illuminating a large area. This procedure is carried out regularly during measurement sessions and for every specimen. An analogue procedure is used for the STEM measurements described later. The diffraction measurements are processed by python-code, where the beam stop is removed from the data, the position of the forward scatter peak is corrected and finally radial intensity distributions are generated.[14] By background and beam modelling the continuous background ("mean scattering") is removed following the method proposed by Tran et al.[16] The remaining oscillations are further Fourier-transformed into the pair distribution function (PDF). The algorithm is shown in more detail in supporting information.

Electron energy loss spectroscopy (EELS) is carried out in a double aberration-corrected JEOL JEM-2200FS equipped with an in-column omega filter and a Gatan Ultrascan 1000XP camera. The electron beam amorphization is initially controlled by EELS in the TEM mode, where control of the diffraction pattern can be carried out more efficiently. Since the ULD position control in TEM mode is extremely difficult due to the surrounding epoxy matrix, the lacey carbon support and the lack of contrast, we additionally utilize STEM-EELS. The measurements are performed at low dose rate conditions of (dose rate of about 3 pA and a semi-convergence angle of about 15 mrad) and at ULD conditions by reducing the microscope field emitter A1 and A2 values to reach an electron flux as low as 0.1 pA. In both cases a suitable region for averaging is chosen in order to mitigate the total dose to a larger amount of material. The critical dose rates for damage-free spectroscopy are estimated as shown in the supporting information (*cf*. Figure S4). The data acquisition and the ULD method are explained in more detail in the supporting information. In all cases, the spectra were processed using the Gatan Microscopy Suite (Gatan, Inc., Pleasanton, CA 94588) and the according modules therein and post-processed using MATLAB (The MathWorks Inc., Natick, Massachusetts, 2020) routines. For every measurement, the spectrometer is aligned and the dispersion is calibrated. The spectra are acquired from a suitable region as spectrum Images that contain the selected spectra at every scan point. This allows for a convenient post-processing and artifact-free controlled averaging. In addition, to the initial low dose core-loss spectra, a low-loss spectrum image was acquired for deconvolution, thickness compensation, and de-scan alignment. After calibration and alignment of the core-loss spectra, background subtraction using the standard "power law" correction at the pre-edge was carried out. These spectra were further deconvolved using the Fourier-ratio method.[8] Further details are shown in the supporting information.

The low-loss spectra were acquired analogously. The plural scattering was removed by Fourier-log deconvolution using the "reflected tail" method that assumes a symmetrical zero-loss peak. The remaining background stems predominantly from the broad plasmon peak. Its shape was modelled by the sum of three gaussians and subtracted effectively extracting the transitions at 3-9 eV.

The energy dispersive X-ray (EDX) spectra were acquired using a JEOL JEM-2200FS double-corrected (S)TEM equipped with a Bruker XFlash 5060 EDX detector. The spectra were generated in scanning mode with a dose rate of about 300 pA for a time of about 15 minutes in order to achieve reasonable spatial distribution (map) as well as a clear spectrum. Further details can be found in the supporting information.

**Molecular dynamics simulations**

Molecular dynamics (MD) simulations were performed using the LAMMPS software package[17,18] in combination with the OPLS-AA force field[19,20] including its extension for halogen bonds.[21] Pre-optimization of individual molecules was performed using the iterative metadynamics with genetic structure crossing (iMTD-GC) algorithm[22,23] using the GFN2-xTB level of theory.[24–26] The pre-optimized molecules were arranged in a lattice with 5.5 Å spacing between the center of mass of each molecule in a 12x12x12 grid (1728 molecules in total) so that they cannot overlap. The resulting structure was then compressed to a density of 1.2 g/cm$^3$ for $AdPh_4$ and 1.8 g/cm$^3$ for $AdBr_3Ph$ using an NVE ensemble and performing MD simulations. Subsequently, several MD simulations were carried out at different temperatures between 300 K and 600 K using an NVT ensemble. The calculation parameters for the different steps are described in the supporting information. This results in a total simulation time of 1.5 ns. For the subsequent electron diffraction simulations with STEMsalabim,[12] the final, fully post-optimized structure was used.

**Density functional theory calculations**

Density functional theory (DFT) calculations in a plane wave-basis were performed with the Vienna ab initio simulation package (VASP).[27,28] The electron−ion interaction was modeled with the projector augmented wave (PAW) approach,[29] whereby the electron−electron exchange and correlation (XC) effects are described in the generalized gradient approximation (GGA)[30] as parametrized in the PBE functional.[31] The electronic wave functions were expanded into plane waves up to a kinetic energy of 410 eV.

The accurate modeling of long-range van-der-Waals interactions is a major challenge for DFT. We account approximately for dispersion interactions using a semi-empirical approach based on the London dispersion formula.[32] Specifically, we use the DFT-D3 suggested by Grimme.[33,34] While this approach is certainly a simple approximation that restricts the predictive power of the calculations with respect to the absolute magnitude of the calculated energy differences, it is expected to describe sufficiently accurately bond lengths and other structural features.

Gas-phase molecule calculations are performed using an orthorhombic 20×20×30 Å³ unit cell. A force threshold of 5 meV/Å is used for structural optimization. The effect of electronic irradiation was modeled by excess electrons, for which a homogeneous background charge was assumed. Optically induced excitonic states were modeled by freezing one electron in the lowest unoccupied molecular orbital (LUMO) and a hole in the highest occupied molecular orbital (HOMO). Spin-conserving and non-spin-conserving transitions were considered.

**Acknowledgement**

We acknowledge support from the German Research Foundation (DFG) in the framework of the research unit FOR2824 (*Amorphous Molecular Materials with Extreme Non-Linear Optical Properties*).

**ASSOCIATED CONTENT**

**Supporting Information**. Supporting information on Optics, Electron diffraction, EELS and EDX, Molecular dynamics and density functional computations as well as organic synthesis is associated with this manuscript.

**Author Contributions**

The manuscript was written through contributions of all authors. All authors have given approval to the final version of the manuscript

**Funding Sources**

German Research Foundation DFG FOR 2824

# Supporting Information

# Adamantanes as white-light emitters: Controlling arrangement and functionality by external Coulomb forces

Jürgen Belz[1], Johannes Haust[1], Marius J. Müller[2], Kevin Eberheim[3], Sebastian Schwan[4], Saravanan Gowrisankar[5], Franziska Hüppe[1], Andreas Beyer[1], Peter R. Schreiner[5], Doreen Mollenhauer[4], Simone Sanna[3], Sangam Chatterjee[2], and Kerstin Volz[1*]

[1] Department of Physics and Materials Science Center, Philipps-University Marburg, Hans-Meerwein Str. 6, 35032 Marburg, Germany.
[2] Institute of Experimental Physics I and Center for Materials Research (ZfM), Justus-Liebig-University Giessen, Heinrich-Buff-Ring, 35392 Giessen, Germany.
[3] Institute of Theoretical Physics and Center for Materials Research (ZfM), Justus-Liebig-University Giessen, Heinrich-Buff-Ring, 35392 Giessen, Germany.
[4] Institute of Physical Chemistry and Center for Materials Research (ZfM), Justus-Liebig-University Giessen, Heinrich-Buff-Ring, 35392 Giessen, Germany.
[5] Institute of Organic Chemistry and Center for Materials Research (ZfM), Justus Liebig University, Heinrich-Buff-Ring 17, 35392 Giessen, Germany.

∗ Corresponding author: kerstin.volz@physik.uni-marburg.de

## I. Optics

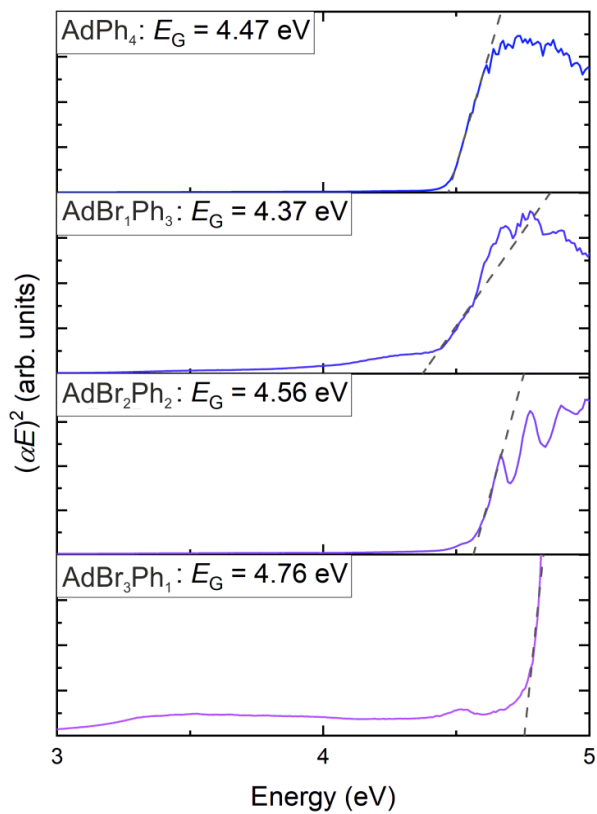

*Figure S1: Tauc plots from the $AdBr_xPh_{4-x}$ family.*

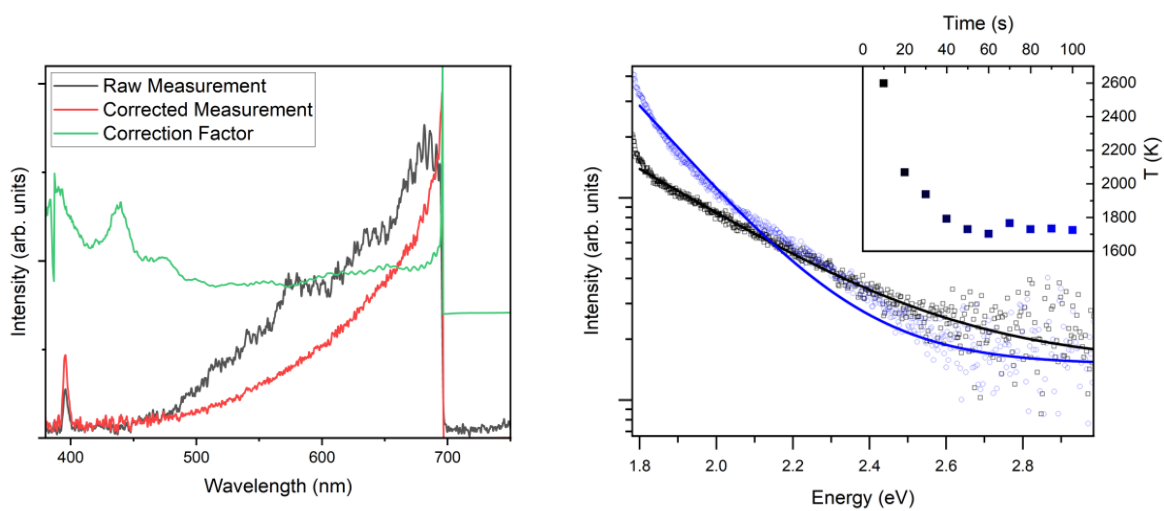

*Figure S2: left: exemplary raw spectral data (black) and intensity calibration curve considering all spectrally non-flat features (optics, lenses, spectrometer, camera etc.) recorded using a traceable incandescent tungsten-halogen source. The corrected spectrum is shown in red. right: exemplary black-body fit curves to the calibrated continuum spectra of $AdBr_2Ph_2$.*

The transition energies are derived from the linear absorption data according to Tauc's method.[1,2] The fit curves and derived transition energies are given as dashed lines together with the respective linear absorption data for the $AdBr_xPh_{4-x}$ family-of-compounds in Figure S1.

The emission data are spectrally corrected using a traceable incandescent tungsten-halogen source placed at the spot of the sample. The as-recorded spectra are corrected by the factor derived from the available reference spectrum after subtraction of the dark spectra, *i.e.*, the data without illumination to account for, *e.g.*, readout noise and dark current; *cf.* left-hand side of Figure S2. The corrected spectra are fit using a black-body curve using the temperature as parameter. Exemplary data are given on the right-hand side of Figure S2. The inset shows the time evolution of the extracted temperature values for constant, continuous illumination.

## II.   Electron diffraction: Pair distribution function and background correction

To describe and analyze the structure of amorphous materials, the so-called reduced pair distribution function G(r) is used. This is a quantity which gives information about the frequency of the next neighbor distances occurring in the material.[3,4] It can be obtained from electron diffraction data through a Fourier transformation of background-corrected intensity. Because experimental electron diffraction data suffers from inelastic and dynamical diffraction, one has to correct the differential diffracted radial intensity *dI* by using an modified empirical model.[5]

$$dI_{corr}(Q) = \frac{\int_{Q_{min}}^{Q_{max}} \overline{f^2(Q')}\, dQ'}{\int_{Q_{min}}^{Q_{max}} \left(\frac{c_0}{Q'^{c_1}} dI_{raw}(Q') - B_N(Q')\right) dQ'} \cdot \left(\frac{c_0}{Q^{c_1}} dI_{raw}(Q) - B_N(Q)\right). \tag{1}$$

Hereby $B_N(Q)$ is a Laurent-type series and Q the scattering vector given as

$$B_N(Q) = \sum_{i=2}^{N} \frac{c_i}{Q^i}, \qquad Q = 4\pi \cdot \frac{sin\left(\frac{2\theta}{2}\right)}{\lambda}. \tag{2}$$

$2\theta$ is electron scattering angle and $\lambda$ the electron´s wavelength. $c_0$, …, $c_N$ are the fitting parameters. The data is fitted by minimizing the quadratic difference $\chi^2$ in the regime of high Q values, called the "tail" of the respective quantity, between $dI_{corr}$ and the mean squared scattering factor $\overline{f^2(Q)}$ of the material

$$\chi^2 = \sum_{tail} \left(dI_{corr}(Q) - \overline{f^2(Q)}\right)^2. \tag{3}$$

The structure factor S(Q) can be then be calculated from $dI_{corr}(Q)$ using the squared mean scattering factor $\overline{f(Q)}^2$ and $\overline{f^2(Q)}$ as

$$S(Q) = 1 + \frac{dI_{corr}(Q) - \overline{f^2(Q)}}{\overline{f(Q)}^2}. \tag{4}$$

Finally, G(r) is obtained by a Fourier transformation of S(Q)

$$G(r) = \int_{Q_{min}}^{Q_{max}} (S(Q) - 1) \cdot Q \cdot sin(Q \cdot r)\, dQ. \tag{5}$$

For complementary simulations, the in-house developed algorithm *STEMsalabim* is used.[6] This is a highly parallel implementation designed for high-performance computing (HPC) clusters and uses the common multi-slice approach.[7] The simulation parameters are chosen in accordance with the experiments and an almost parallel illumination, which has been shown is the most suitable case for such amorphous materials,[3] are applied. For these small semi-convergence angles, lens aberrations of the microscope do not affect the structure factor obtained from the simulation data for amorphous materials.[8] Accordingly, the aberrations are not considered here. Since per construction inelastic scattering is suppressed in the simulations, the previously described background correction in equation (1) is not necessary and S(Q) can be directly calculated out of $dI_{raw}(Q)$ using equation (4).

The input for these electron diffraction simulations are amorphous $AdPh_4$ and $AdBr_3Ph$ cells, which were generated by molecular dynamics simulations, as described in the following section.

Experimental scattering data of $AdPh_4$ is shown in Figure S3 (left) for the laser-irradiated and (right) the electron-irradiated state. Both diffraction patterns show the same basic ring structure.

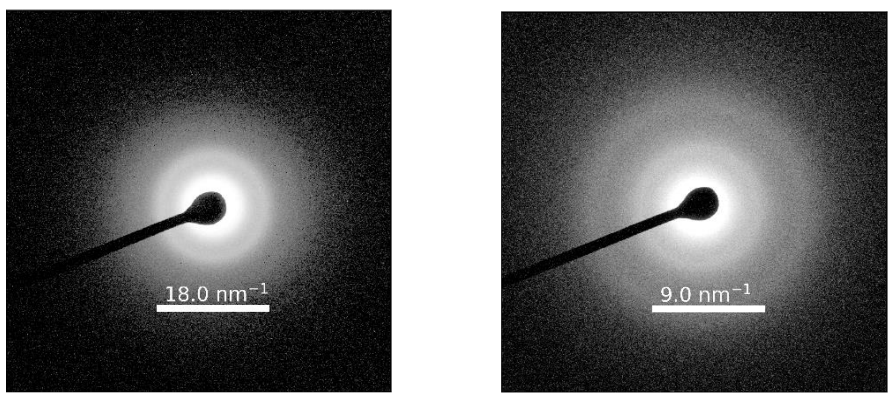

*Figure S3: left: Diffraction pattern as recorded from previously laser irradiated $AdPh_4$ right: The diffraction pattern of an electron-irradiated specimen shows the same basic ring pattern.*

### III.  Molecular dynamics simulation

Different relaxation and simulation steps of the MD simulation:

1. Compression to 1.2 g/cm³ and 1.8 g/cm³ (20000 steps at 0.01 fs, NVE)

2. Simulation at 400 K (10000 steps at 0.1 fs, NVT)

3. Simulation at 400 K (10000 steps at 0.5 fs, NVT)

4. Simulation at 600 K (1000000 steps at 0.5 fs, NVT)

5. Simulated annealing from 600 K to 300 K (1000000 steps at 0.5 fs, NVT)

6. Simulation at 300 K (1000000 steps at 0.5 fs, NVT)

## IV. EELS measurements and dose rate estimations

In order to achieve ultra-low dose (ULD) conditions, we reduced the emission current of the field emitter from around 135 µA to 95 µA which led to a reduction of the minimal electron dose rate of about 3 pA to 0.15 pA. In addition, the reduction of the anode voltages leads to an improvement of the FWHM of the direct beam energy distribution measured by the zero-loss peak from 1.2 eV to about 0.7 eV. Due to emphasis on dose efficiency, a collection semi-angle of about 9 mrad was chosen which practically limited the energy resolution to about 1.0 eV. From diffraction studies, we derived a limit of an acceptable dose that is needed for "artefact-free" measurements of the pristine state, i.e., ~20 e/Å². Due to the high electron flux per area in aberration corrected scanning transmission electron microscopes (STEM), the total dose per unit area has to be tailored carefully in order to keep the collected signal above the noise level of the CCD camera as well as the total dose close to the maximum acceptable value. Hereto, we estimate the defocus (~3 microns) needed for a given dose rate (e.g., 3 pA) and a given pixel time (0.5 s). In addition, we estimate the resolution (i.e., the resulting illuminated area) and the optimum scan density for non-overlapping probes (80x80 nm²). These approximations are based on a simple trigonometric picture of the diverging probe with a given convergence semi-angle (e.g., 21 mrad).

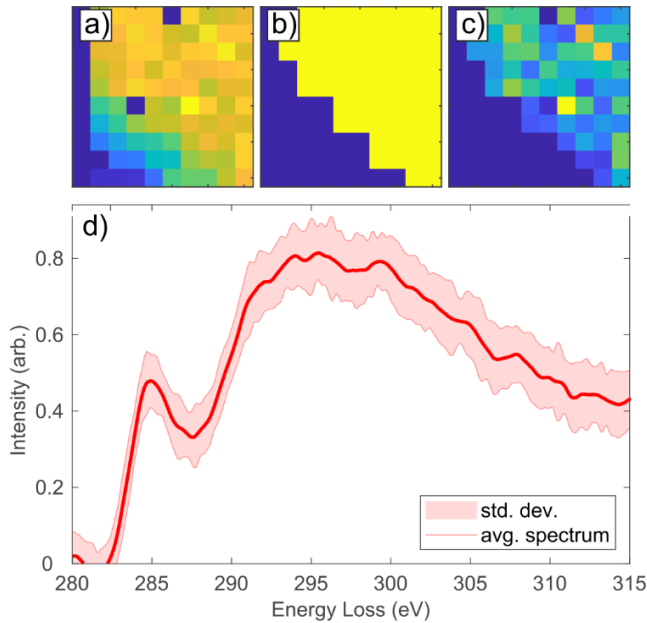

*Figure S4: (a) Spectrum image showing summed intensity loss. The recorded map can be cropped (b) and outlier corrections applied. The resulting sanitized map (c) leads to a representative averaged spectrum for a given dataset (d).*

With these estimations, we locate the specimen regions with a minimum dose rate. Using the *Gatan Microscopy Suite*, we acquire so-called spectrum images that match the calculated step size and therefore have non-overlapping pixels / probe positions (Figure S4a). After each core-loss map, a low-loss map was acquired in order to correct the plural scattering broadening by means of Fourier-ratio deconvolution.[9] In addition, the individual spectra of a map were aligned and calibrated. Finally, a *MATLAB* based routine removed outliers in the datasets (X-rays) and a representative area is selected for summation (Figure S4b,c). By doing so, we can avoid unwanted overlap with epoxy or lacey carbon and can sum up regions of similar thicknesses. The resulting summed spectra are furthermore summed (averaged) for different positions/grains (Figure S4d).

## V. EELS and EDX of $AdBr_3Ph$

The element composition of the specimen can be determined, within limits, by analytical micro characterization techniques. Firstly, the characteristic X-ray emission spectrum induced by high energy electron irradiation can be measured by EDX (Energy dispersive X-ray spectroscopy) in the transmission electron microscope (TEM). Secondly, the energy loss of the primary electrons can be measured by EELS (Electron energy loss spectroscopy).

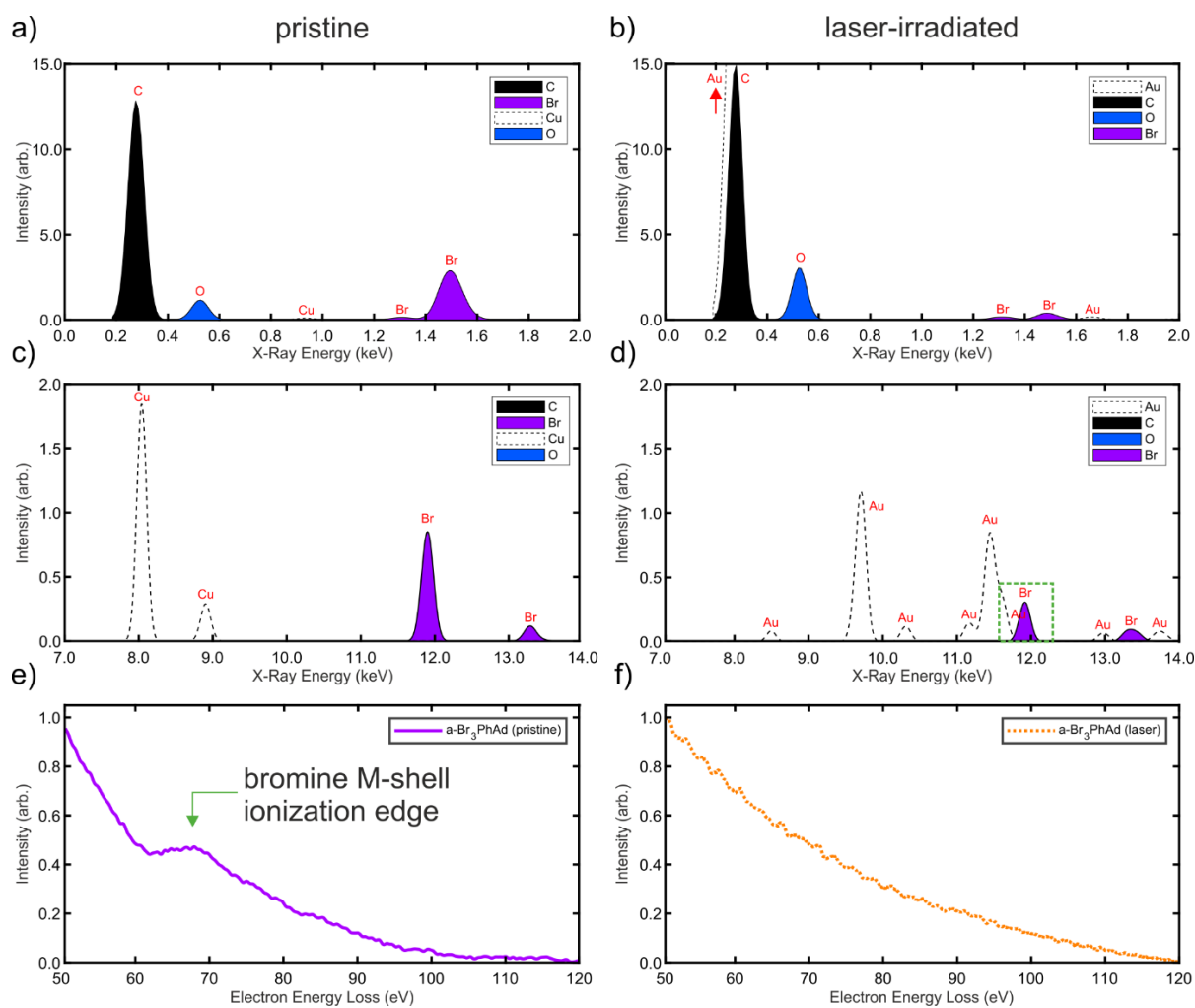

*Figure S5: (a,c,e) Show the analytical data derived from pristine $AdBr_3Ph$ on a copper support grid, whereas (b), (c), and (f) show the spectra from the laser-irradiated specimen on a gold support grid. The EDX spectra (a-d) are normalized to the integrated carbon signal, and suggest the absence of significant amounts of bromine in the case of laser irradiation (c,d). The bromine peaks shown in (d) have a strong overlap with the geometrical background signal (dashed lines) from the gold support grid (green box, dashed) that leads to the false interpretation of bromine incorporation. The additional EELS measurements clearly show the existence of bromine in the pristine specimen (e) and the absence of bromine in the laser-irradiated specimen (f).*

Figure S5a,c and e) show the analytical data for $a$-$AdBr_3Ph$ (pristine). The EDX spectra Figure S5a-d are normalized to the integrated carbon signal, and shown after continuum background correction, "series-fit" deconvolution using theoretical Cliff-Lorimer factors.[10] All steps were carried out using the *Bruker* Software *Suite Esprit 2.3*.

In the case of the unirradiated specimen, one can see bromine K-shell (a) as well as L-shell emission (b) of X-rays, clearly indicating the existence of bromine. It is noteworthy that Figure S5b also shows a strong geometrical background signal from the surrounding copper grid, which is inevitable due to high energy backscattering of 200 kV electrons. These artifacts are plotted as dash-dotted lines for reference only.

Since EDX can produce these unwanted geometrical artifacts, we verified the presence of bromine by measuring the M-shell ionization edge by EELS (f). In contrast to the findings in the pristine specimen, we only find a very small bromine L-shell signal (b) and K-shell peak (d) for the laser-irradiated specimen. Since the signal itself is small already, it cannot be fully distinguished from the gold artifact (d), since the sub-shell lines coincide almost exactly. By using EELS (f), which is free of geometrical artifacts and generally more sensitive as well as dose efficient than EDX,[9] we can verify that the bromine amount is at most below 1 at%. In combination with the finding of Figure 7a and Figure 8b in the main manuscript, we can conclude that the amount of bromine is close to zero.

## VI. *Density functional theory computations*

In the following, the impact of electron and laser irradiation on single $AdBr_xPh_{4-x}$ clusters are discussed in detail. Table S1 reports the calculated C-Br distance as a function of the excess electrons and electron-hole pairs. The C-Br distance in $AdBr_4$ in the neutral ground state is 1.99 Å. For distances larger than 2.50 Å, Br is considered as detached form the adamantane core.

|  | -2 el. | -1 el. | neutral | +1 el. | +2 el. | +3 el. | +4 el. | +1 e-h | +2 e-h |
|---|---|---|---|---|---|---|---|---|---|
| Ad Ph$_4$ | --- | --- | --- | --- | --- | --- | --- | --- | --- |
| Ad Br$_1$Ph$_3$ | 1.98 Å | 1.99 Å | 2.01 Å | > 2.5 Å | > 2.5 Å | > 2.5 Å | > 2.5 Å | 2.02 Å | 2.02 Å |
| Ad Br$_2$Ph$_2$ | 1.97 Å | 1.98 Å | 2.00 Å | 2.32 Å | > 2.5 Å | > 2.5 Å | > 2.5 Å | 2.01 Å | 2.01 Å |
| Ad Br$_3$Ph$_1$ | 1.97 Å | 1.98 Å | 2.00 Å | 2.23 Å | > 2.5 Å | > 2.5 Å | > 2.5 Å | 2.00 Å | > 2.5 Å |
| Ad Br$_4$ | 1.96 Å | 1.97 Å | 1.99 Å | 2.19 Å | 2.33 Å | > 2.5 Å | > 2.5 Å | 2.17 Å | 1.98 Å (2x) <br> > 2.5 Å (2x) |

*Table S1: Calculated C-Br distance as a function of the excess electrons and electron-hole pairs.*

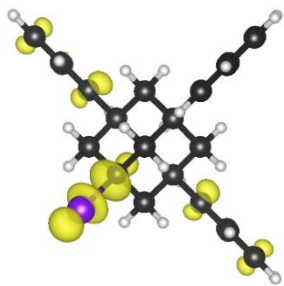

Figure S6: LUMO of the $AdBr_3Ph_1$ molecule as calculated by DFT with a plane wave basis. The isosurfaces 0.007 eV/Å³ are shown.

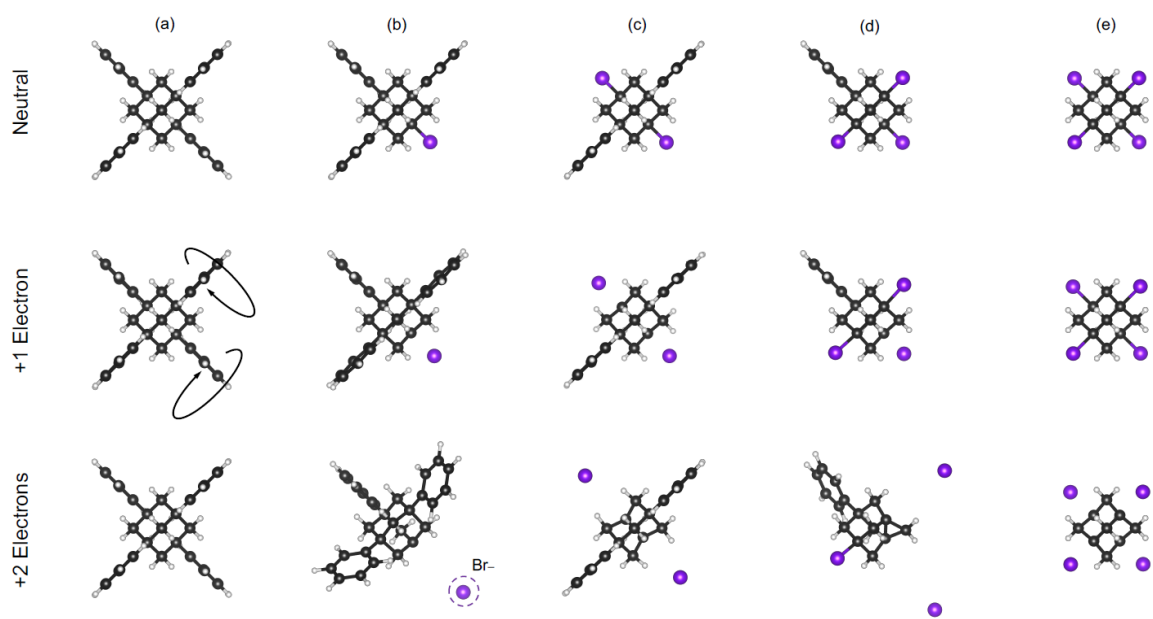

Figure S7: Structural models of the $AdBr_xPh_{4-x}$ series in the neutral charge state (first row) as well as upon electron doping. The cluster structure is calculated within DFT-PBE and a plane-wave basis.

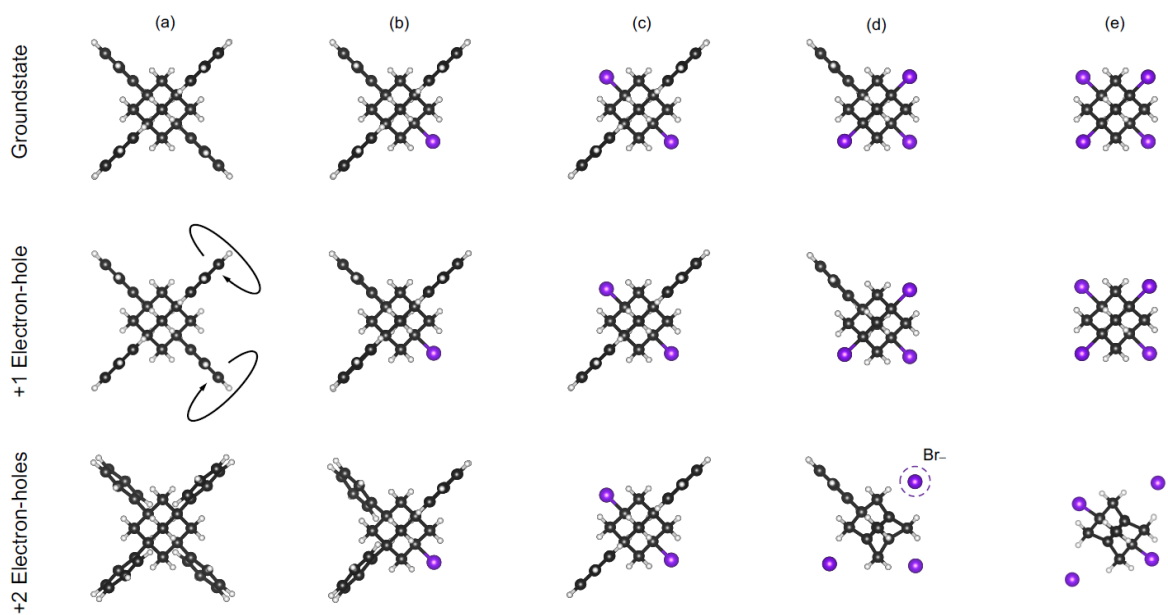

Figure S8: Structural models of the $AdBr_xPh_{4-x}$ series in the electronic ground state (first row) as well as formation of one or more electron-hole pairs. The cluster structure is calculated within DFT-PBE and a plane-wave basis.

## VII. Organic synthesis: experimental procedures

*1,3,5,7-Tetrabromoadamantane (1)*[11,12]

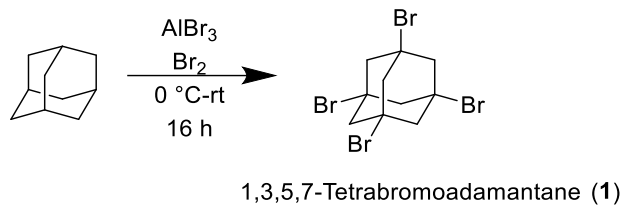

1,3,5,7-Tetrabromoadamantane (**1**)

A mixture of anhydrous AlBr$_3$ (1.3 g, 0.05 mol) and 3 mL bromine was cooled to 0 °C. Adamantane (1.5 g, 0.011 mol) was added in small portions through a solid addition funnel. The mixture was then allowed at rt for 16 h. HBr gas that evolved was passed through a second flask containing concentrated aqueous sodium bisulfite. When no more HBr evolved, the reaction mixture was treated first with 20 mL saturated sodium bisulfite solution and then with 20 mL 6N HCl. The resulting dark yellow precipitate was taken up in chloroform, washed with distilled water and brine, and dried over anhydrous magnesium sulfate. The organic solution was filtered and evaporated at reduced pressure; the residue was dissolved from acetone/water to afford 2.2 g (45%) of **1** as a white solid. The white solid was dissolved in CHCl$_3$/methanol (8:2) gave pure title compound.

**$^1$H NMR** (400 MHz, CDCl$_3$) δ = 2.58 (s, 12H).

**$^{13}$C NMR** (101 MHz, CDCl$_3$) δ = 54.83, 54.62.

**Elemental analysis** calcd (%) for C$_{10}$H$_{12}$Br$_4$: C 26.58, H 2.68; found: C 26.76, H 2.66.

**Note:** In the presence of AlCl$_3$, **1** was obtained along a trace amount of 1-chloro-3,5,7-tribromoadamantane.

*Synthesis of 1,3,5-tribromo-7-phenyladamantane (2)*[13,14]

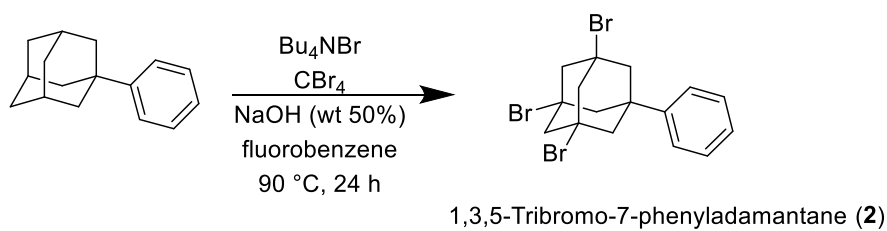

1,3,5-Tribromo-7-phenyladamantane (**2**)

To a solution of 1-phenyladamantane (0.42 g, 2.0 mmol), CBr$_4$ (5.3 g, 8.0 mmol), and Bu$_4$NBr (0.066 g, 0.20 mmol) in 6 mL of fluorobenzene was added 6 mL NaOH (wt 50%) and the resulting reaction mixture was heated to 90 °C for 24 h. The solvent was evaporated under reduced pressure. The crude product was suspended in water and extracted three times with 20 mL of DCM. The combined organic layers were dried over Na$_2$SO$_4$, filtered, and the solvent was evaporated under reduced pressure to give **2** as a solid (0.6 g, 66%). The solid was additionally purified via sublimation.

**$^1$H NMR** (400 MHz, CDCl$_3$) δ = 2.38 (s, 6H), 2.58 (s, 6H), 7.25 (m, 5H).

**$^{13}$C NMR** (101 MHz, CDCl$_3$) δ = 46.37, 50.63, 56.07, 57.68, 124.51, 127.24, 128.82, 144.48.

**Elemental analysis** calcd (%) for C$_{16}$H$_{17}$Br$_3$: C 42.80, H 3.82; found: C 42.77, H 3.82.

*Synthesis of 1,3-dibromo-5,7-diphenyladamantane (3)*

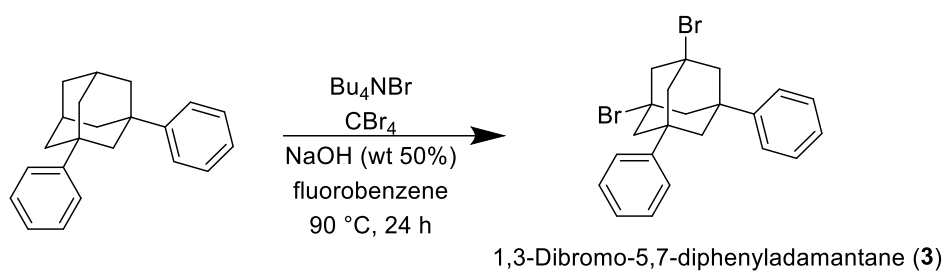

1,3-Dibromo-5,7-diphenyladamantane (**3**)

To a solution of 1,2-diphenyladamantane (0.3 g, 1.04 mmol), CBr$_4$ (2.6 g, 8.0 mmol), and Bu$_4$NBr (0.03 g, 0.10 mmol) in 3 mL of fluorobenzene was added 6 mL NaOH (wt 50%) and the resulting reaction mixture was heated to 90 °C for 24 h. The solvent was evaporated under reduced pressure; the crude product was suspended in water and extracted three times with 20 mL of DCM. The combined organic layers were dried over Na$_2$SO$_4$ and filtered and the solvent was evaporated under reduced pressure to give the **3** as a solid (0.27 g, 61%). The solid was additionally purified by column chromatography (Hexane:EA, 95:5).

**$^1$H NMR** (400 MHz, CDCl$_3$) δ = 2.13 (s, 2H), 2.58 (s, 8H), 2.97 (s, 2H), 7.37 (m, 10H).

**$^{13}$C NMR** (101 MHz, CDCl$_3$) δ = 44.43, 45.85, 51.64, 57.37, 61.15, 124.68, 126.92, 128.69, 146.03.

*Synthesis of 1-bromo-3,5,7-triphenyladamantane (4)*[15–17]

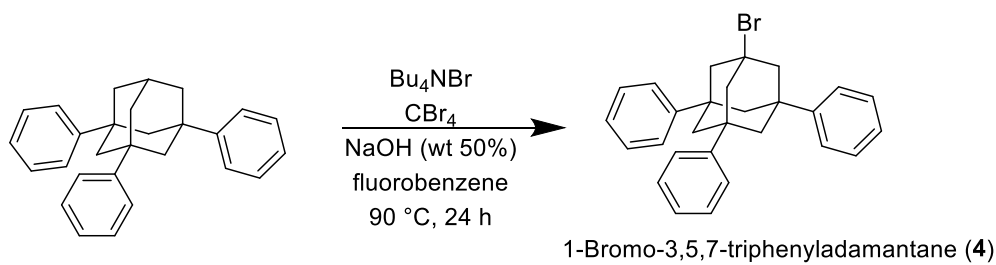

To a solution of 1,3,5-triphenyladamantane (0.40 g, 1.09 mmol), CBr$_4$ (1.46 g, 4.4 mmol), and Bu$_4$NBr (0.04 g, 0.12 mmol) in 3 mL of fluorobenzene was added 6 mL NaOH (wt 50%) and the resulting reaction mixture was heated to 90 °C for 24 h. The solvent was evaporated uner reduced pressure. The crude product was suspended in water and extracted three times with 20 mL of DCM. The combined organic layers were dried over Na$_2$SO$_4$ and filtered and the solvent was evaporated under reduced pressure to give **4** as a yellow solid (0.37 g, 75%). The yellow solid was additionally purified via sublimation.

**$^1$H NMR** (400 MHz, CDCl$_3$) δ = 2.38 (s, 6H), 2.75 (s, 6H), 7.19 (m, 5H), 7.29 (m, 5H).

**$^{13}$C NMR** (101 MHz, CDCl$_3$) δ = 42.13, 46.58, 52.69, 65.14, 125.08, 126.57, 129.07, 147.67.

**Elemental analysis** calcd (%) for C$_{28}$H$_{27}$Br: C 75.84, H 6.14; found: C 76.07, H 6.09.

*Synthesis of 1,3,5,7-tetraphenyladamantane (5)*

The synthesis of **5** has been reported multiple times and followed the initial protocol of Newman.[15] All spectral data agree with the literature.

  **NMR spectral data of all synthesized compounds**

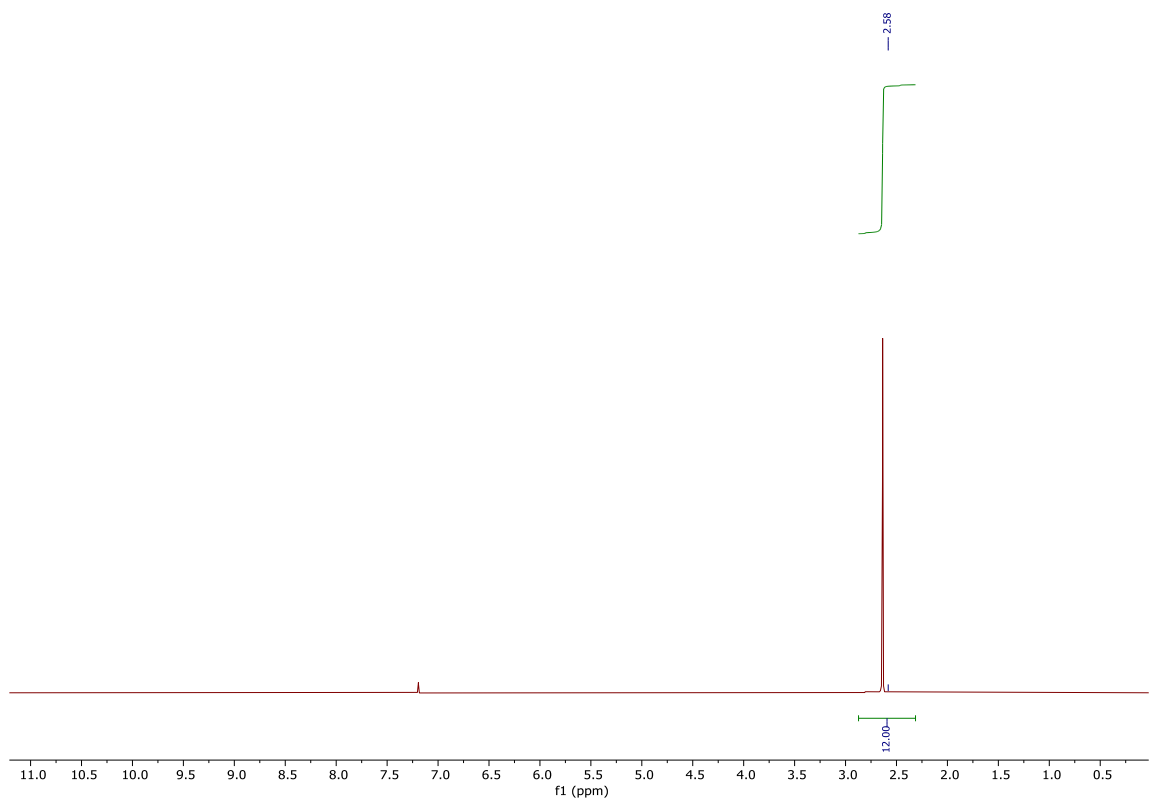

*Figure S6: ¹H NMR spectrum of 1,3,5,7-tetrabromoadamantane (**1**).*

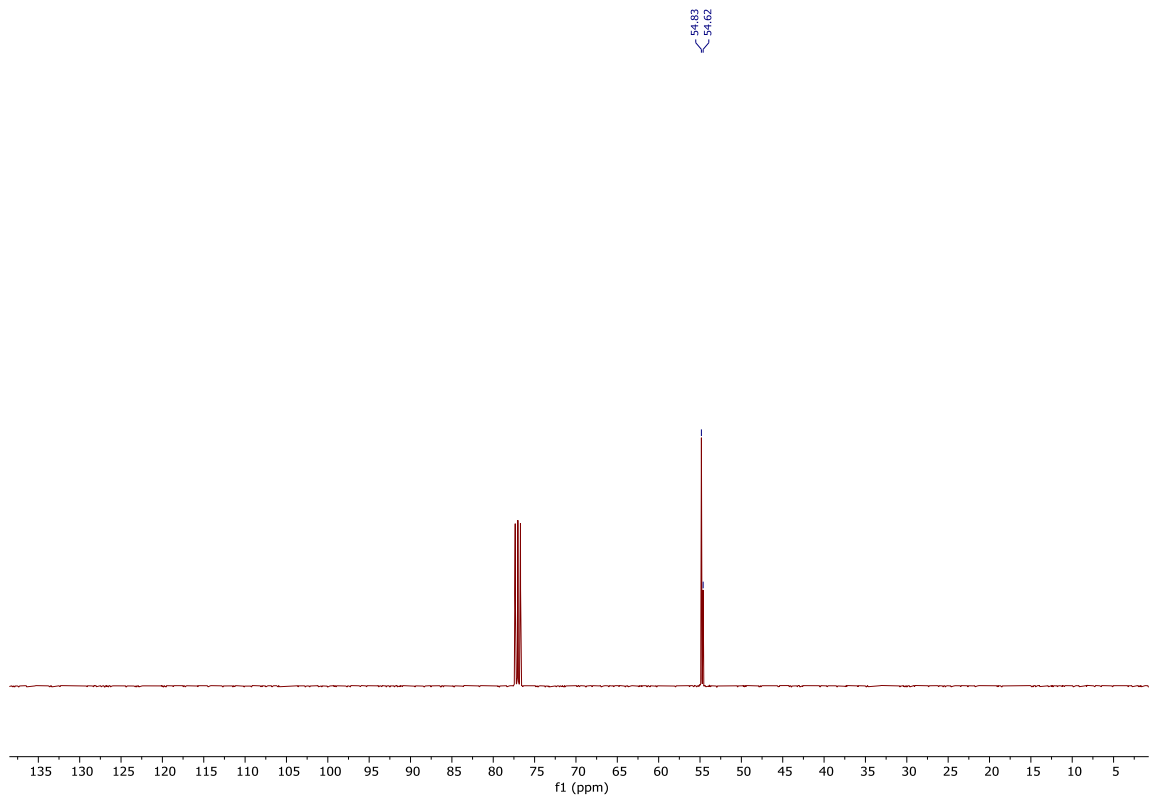

*Figure S7: 13C NMR spectrum of 1,3,5,7-tetrabromoadamantane (1).*

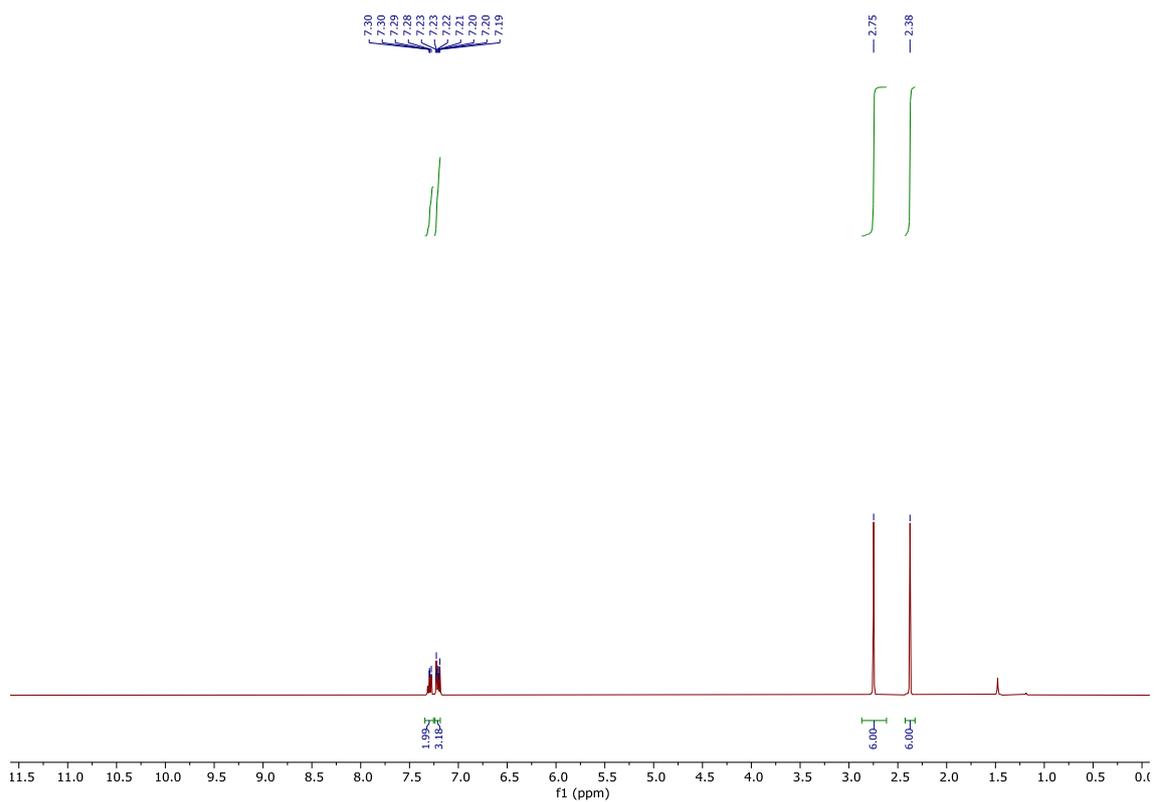

*Figure S8: 1H NMR spectrum of 1,3,5-tribromo-7-phenyladamantane (2).*

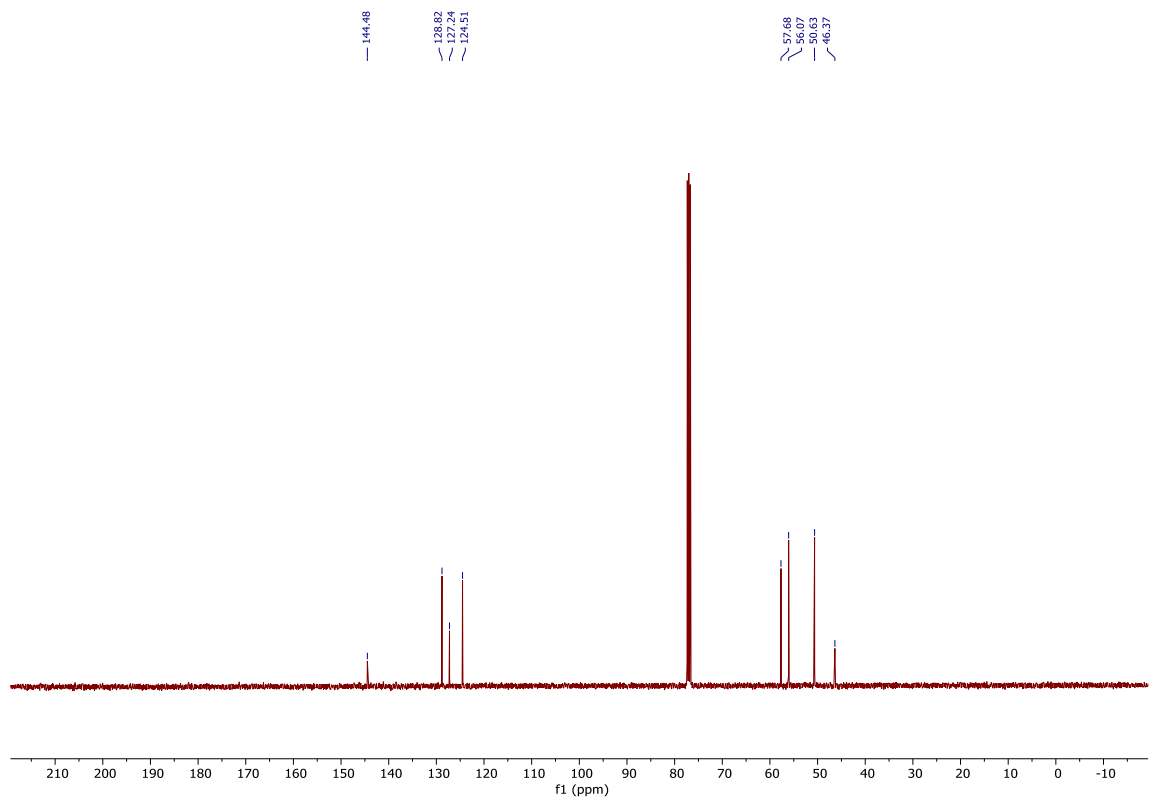

*Figure S9: 13C NMR spectrum of 1,3,5-tribromo-7-phenyladamantane (2).*

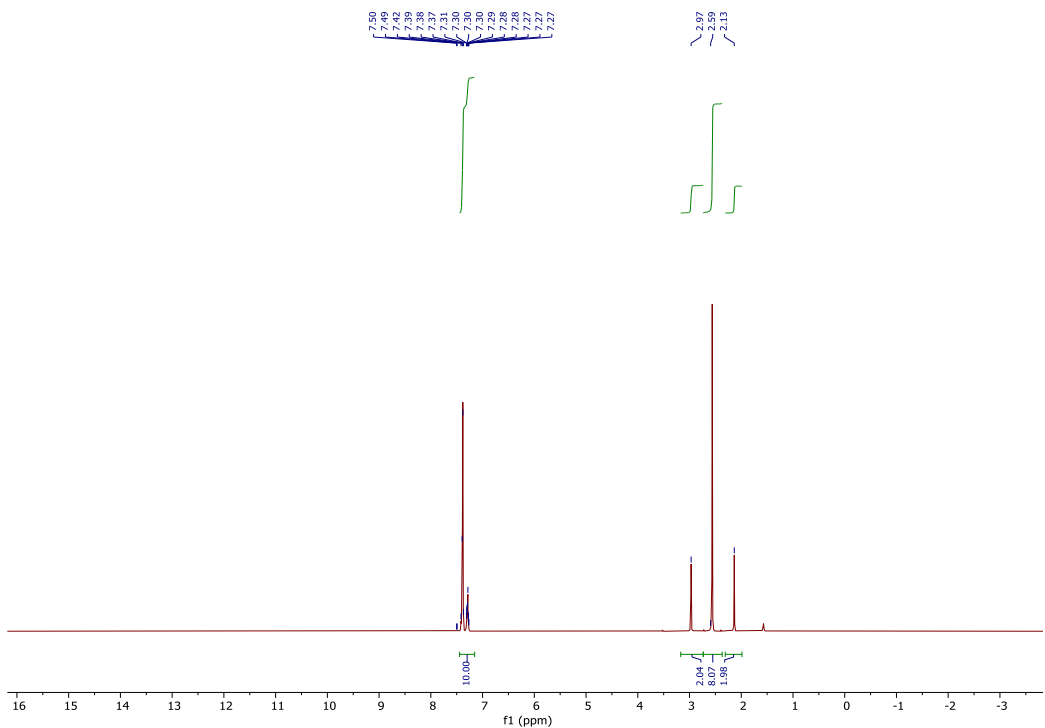

*Figure S10: 1H NMR spectrum of 1,3-dibromo-5,7-diphenyladamantane (3).*

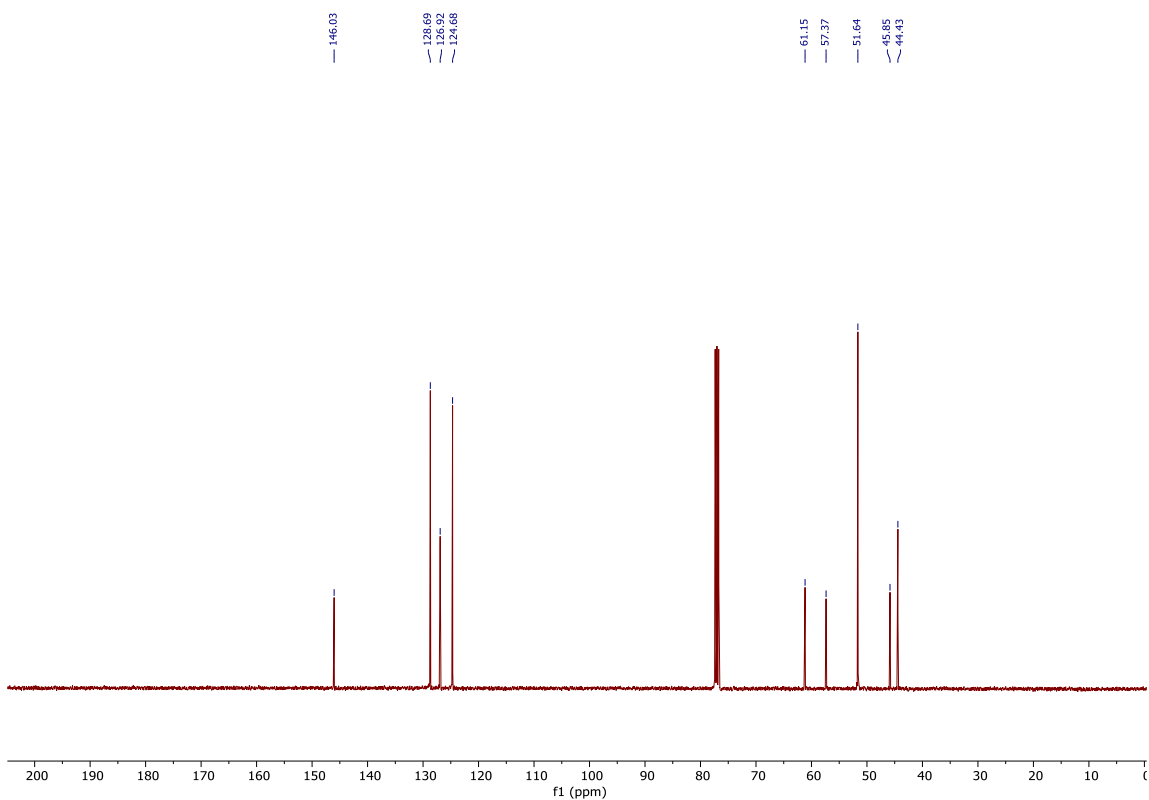

*Figure S11: 13C NMR spectrum of 1,3-dibromo-5,7-diphenyladamantane (3).*

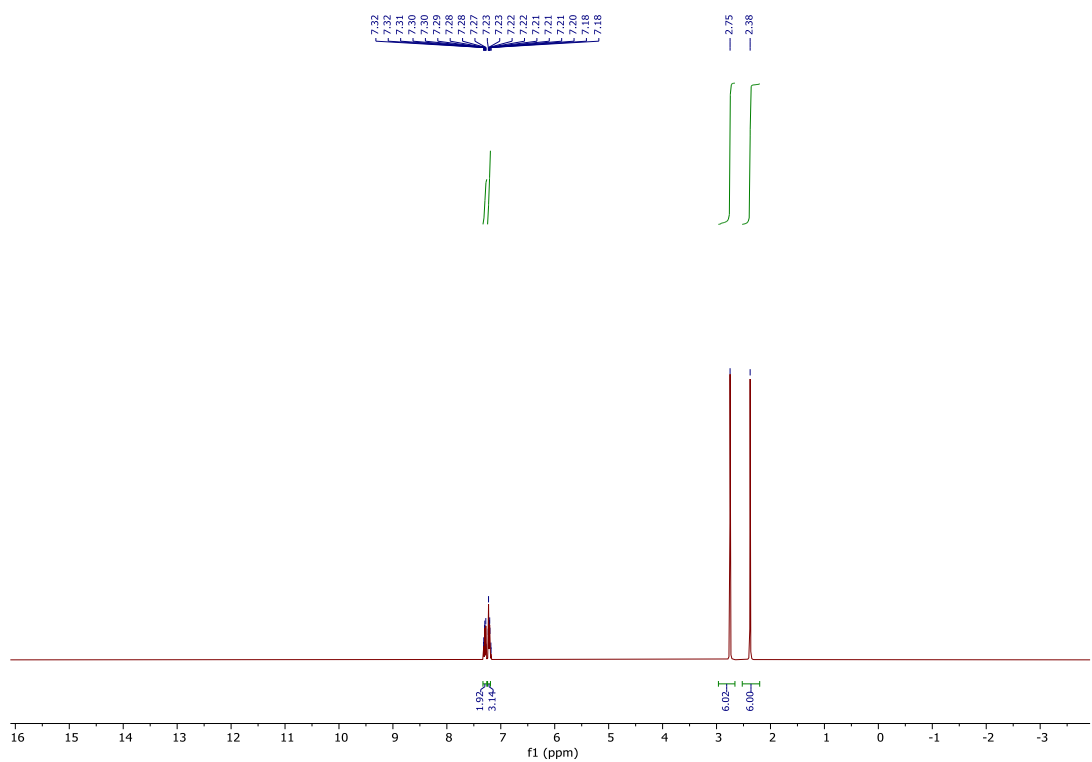

*Figure S12: 1H NMR spectrum of 1-bromo-3,5,7-triphenyladamantane (4).*

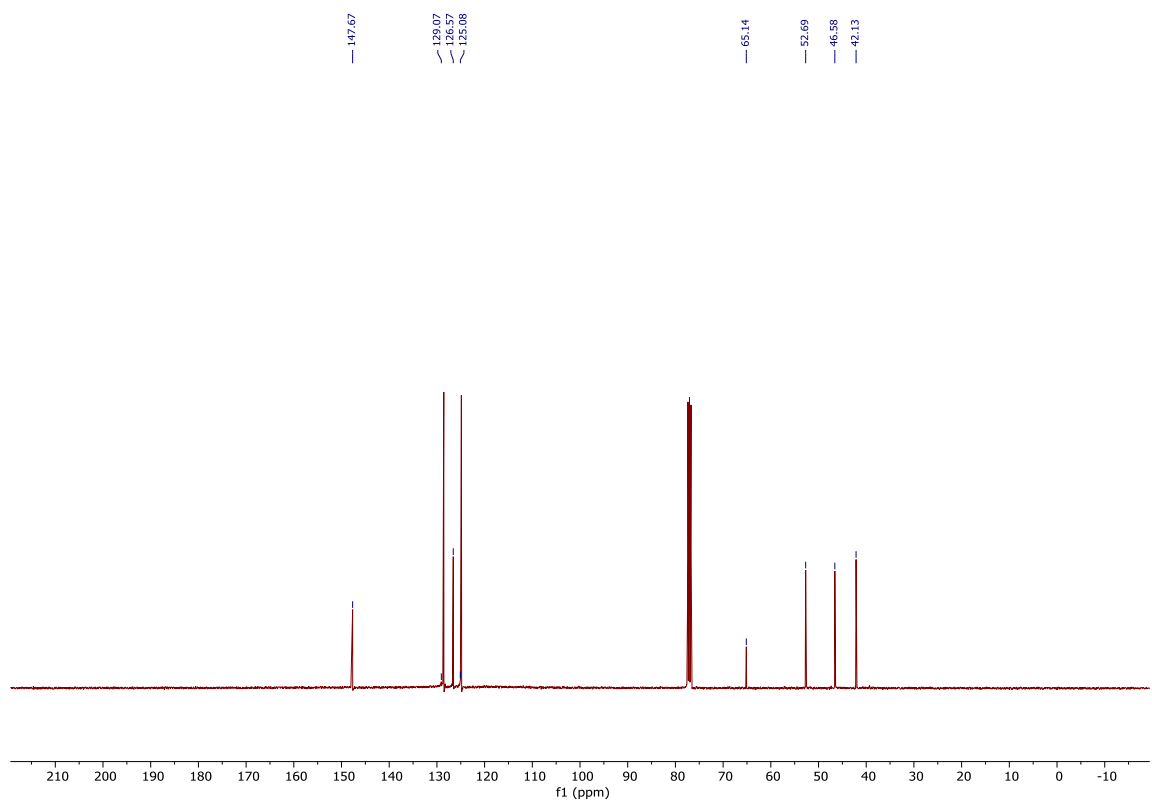

*Figure S13: 13C NMR spectrum of 1-bromo-3,5,7-triphenyladamantane (4).*

**SI References**

Substituents. *Tetrahedron* **2013**, *69* (15), 3238–3248. https://doi.org/10.1016/j.tet.2013.02.042.

(15) Newman, H. Facile Syntheses of 1,3-Diphenyl-, 1,3,5-Triphenyl-, and 1,3,5,7-Tetraphenyladamantane from 1-Bromoadamantane. *Synthesis (Stuttg).* **1972**, *1972* (12), 692–693. https://doi.org/10.1055/s-1972-21969.

(16) Fokin, A. A.; Schreiner, P. R. Metal-Free, Selective Alkane Functionalizations. *Adv. Synth. Catal.* **2003**, *345* (910), 1035–1052. https://doi.org/10.1002/adsc.200303049.

(17) Nasr, K.; Pannier, N.; Frangioni, J. V.; Maison, W. Rigid Multivalent Scaffolds Based on Adamantane. *J. Org. Chem.* **2008**, *73* (3), 1056–1060. https://doi.org/10.1021/jo702310g.